\documentclass[twocolumn]{aastex63}

\usepackage{bm}
\usepackage{wasysym}
\usepackage{xcolor}
\usepackage{soul}

\submitjournal{ApJ}
\shorttitle{Strong stellar winds interacting with a SMBH}
\shortauthors{Lora et al.}

\begin{document}
\title{Huffing, and puffing, and blowing your house in: \\
Strong stellar winds interaction with a super massive black hole}
%

\author[0000-0003-3588-5235]{V. Lora}
\affiliation{Instituto de Radioastronom{\'i}a y Astrof{\'i}sica, Universidad Nacional Aut\'onoma de M\'exico,\\
Apartado Postal 72-3 (Xangari), Morelia, Michoc{\'a}n 58089, Mexico}

\author[0000-0002-0835-1126]{A. C. Raga}
\affiliation{Instituto de Ciencias Nucleares, Universidad Nacional Aut\'onoma de M\'exico,\\
Apartado Postal 70-543, 04510 Ciudad de M\'{e}xico, Mexico}

\author[0000-0003-3863-7114]{J. Cant\'o}
\affiliation{Universidad Nacional Autónoma de México, Instituto de Astronomía, AP 70-264, CDMX 04510, México}

\author[0000-0001-7222-1492]{A. Esquivel}
\affiliation{Instituto de Ciencias Nucleares, Universidad Nacional Aut\'onoma de M\'exico,\\
Apartado Postal 70-543, 04510 Ciudad de M\'{e}xico, Mexico}


\email{v.lora@irya.unam.mx}

\begin{abstract}
We present analytic and numerical models of a cluster wind flow resulting from the interaction of stellar winds of massive stars, with a super massive black hole (SMBH). We consider the motion of the stars as well as the gravitational force of the SMBH. In the numerical simulations we consider two cases: the first one with the stars is in circular orbits, and the second one with the stars in eccentric orbits around the SMBH. We found that after the system reaches an equilibrium, the circular and elliptical cases are very similar.
We found a very good agreement between the analytical and numerical results, not only from our numerical simulations but also from other high resolution numerical calculations. The analytical models are very interesting, since the properties of such complex systems involving strong winds and a massive compact object, can be rapidly inferred without the need of a numerical calculation.

\end{abstract}

\keywords{
methods: hydrodynamics, numerical --— galaxies: black holes --- stars: winds; Wolf-Rayet}
%

\section{Introduction}
\label{sec:intro}

The volume of a sufficiently evolved stellar cluster is filled by the
winds from the cluster stars. The interacting stellar winds produce an
inhomogeneous, outwards flow which has been called the ``cluster wind''.

This cluster wind was described in terms of a steady, mass loaded flow, analytically by \cite{can00}. In an earlier paper, \citet{chevalier:85} found an equivalent solution for a cluster wind driven by supernova explosions.

The mass loaded cluster wind model was explored in more detail by
\cite{silich:04, rodg07} and \cite{pal13}
who studied the effects of different stellar
distributions (within the cluster) and radiative energy losses.
\cite{falle02} used the same ``cluster wind solution'' to model
the flow produced by a group of photoevaporating gas clumps.

Full, 3D gasdynamical simulations of the production of a cluster
wind through the interaction of many stellar winds were presented by
\cite{rag01,rodg07,rodg08,hueyotl:10,pal13}. The related problem of the detonation of
a single supernova within a cluster wind was studied by \cite{rodr14}
and \cite{cas:15}.

None of these studies considered either the motions of the stars within the cluster nor the effect of the cluster's gravitational field. This is justified since the velocities of the stars (typically of a few km~s$^{-1}$) and the escape velocity from cluster are much lower than the stellar wind velocities (of $\sim 10^3$~km~s$^{-1}$).

An interesting case is where the orbital velocities of the stars in a cluster are comparable with the velocity of the stellar winds, for instance if the stellar cluster contains in its center a super massive black hole (SMBH) (or of intermediate mass). In such a cluster the orbital velocities of the stars could be as high as $\sim10^{4}$~km~s$^{-1}$.

A clear example of this situation are the stars orbiting the black hole at the centre of the Galaxy (see, e.g., \citeauthor{gen03} \citeyear{gen03}, \citeauthor{ghez:05} \citeyear{ghez:05}), associated with the Sgr~A$^{*}$ radio source.

\cite{yusef:16} pointed out that the extra kinetic energy of the stellar winds resulting from the rapid stellar motions, can help to explain the high $\sim 4\times 10^7$~K temperature of the Galactic centre X-ray emitting bubble \citep{bag03,wang13}, because the stellar wind velocities (of $\sim 1000$~km~s$^{-1}$) alone are definitely insufficient to produce the observed temperature.

Inspired by the observations  we study cluster wind models for a system of stars that are orbiting around a central, massive black hole. In our models we consider both the motion of the stars and the action of the gravitational force (from the central black hole) on the cluster wind.

Several papers have explored numerical models of cluster winds and a SMBH (Sgr A*)
\citep{rockefeller:04, cuadra:05, cuadra:06, cuadra:08, lutzgendorf:16}.
\cite{ressler:18} model the winds from 30 Wolf-Rayet stars that dominate the accetion budget in Sgr A*. They include the radiative cooling, collisional ionization equilibrium,up-to-date stellar mass-loss rates, wind velocities and locations of the closest stars of Sgr A*. Very recently \cite{ressler:20} performed MHD simulations of Sgr A* and magnetized winds of Wolf-Rayet stars orbiting it. They found a very small impact of magnetic fields in the accretion of material in Sgr A* from Wolf-Rayet stellar winds.

In the present paper we do not take into account magnetic fields.
We approach the problem in two ways:
\begin{itemize}
\item a generalization of the analytical steady, mass loaded flow solution of \cite{can00}
  to include the stellar motions and the gravitational force of the black hole,
\item 3D gas-dynamic simulations including the winds from $\sim 100$ orbiting
  stars for two cases: circular and eccentric orbits.
\end{itemize}
The analytic models that we develop are similar to the
ones of \cite{silich:08}, who also modeled
a wind from a cluster with a central SMBH. Our models
differ from their results in that we include both
the gravitational pull of the SMBH and the motions
of the stars. These motions were not included in
the analytic model of \cite{silich:08}.

Our analytic models are based on equations similar to the ones
of previous papers:
\begin{itemize}
\item \cite{qua:04} derived the gasdynamic equations
  for a mass-loaded wind from a cluster with a central, massive
  compact object, and carried out time-integrations to obtain
  the steady, critical wind solution,
\item \cite{silich:08} presented analytical considerations and
  numerical solutions of the steady state version of the same
  equations,
\item \cite{shc:10} presented numerical integrations of the cluster
  wind+central massive object problem including a thermal conduction
  and a two-temperature description of the flow,
\item \cite{yal:18} studied the same problem as \cite{silich:08},
  and presented analytic solutions for different limiting regimes.
\end{itemize}
The models presented in these papers differ from ours in that
they do not include the motion of the stellar wind sources when
calculating the energy injected by the stellar winds into the
cluster wind.

Our models do not consider radiative cooling for the cluster
wind flow (which is appropriate for the Galactic centre cluster).

The paper is organized as follows. In Section 2 we develop the mass loaded wind formalism, obtain solutions and explore different flow parameters. In Section 3 we present a full, 3D gas-dynamic simulation of the cluster wind flow. Finally, we discuss our results in Section 4, and the provide with the conclusions in Section 5.

\section{The cluster wind as a steady, mass-loaded flow}
\label{sec:analytic}
\subsection{General considerations}
We consider a stellar cluster with a central massive object of mass $M_{bh}$. We assume that the stars within the cluster have identical stellar winds, with a mass loss rate ${\dot M}_w$ and terminal velocity $v_w$, and that the production of these winds is not affected by the possible near presence of the massive central object. In addition, we assume that the stars have a uniform distribution (with a number density  $n$ of stars per unit volume),  inside an outer cluster radius $R_c$, and that the cluster has many stars, so that the ``cluster wind'' (produced by the merging of the winds from the individual stars) can be modeled with a ``mass loaded flow'' formalism, in which the stellar winds are included as a continuous source of mass and energy. Furthermore, we consider a time-independent configuration of a steady cluster wind flow.

The rate of mass injection (per unit volume and time) is:
\begin{equation}
  {\dot m}=n{\dot M}_w\,,
  \label{min}
\end{equation}
which is independent of position in our constant $n$ and ${\dot M}_w$ cluster.

If the stars are moving in random orbits around the central black hole, the net injection of linear and angular momentum due to the orbital motion of the stellar wind sources is zero. This of course is true only in the limit of a high
  spatial density of stellar wind sources. For a real case in which the
  number of cluster stars is not so large, localized regions of organized
  linear and angular momentum are likely to exist, and they will
  drive turbulence in the cluster wind. To describe this effect,
  one has to go beyond the simple, mass-loading formalism
  which we are using here, and we will therefore assume no net momentum
  injection from the stellar winds.


In order to calculate the kinetic energy injection from the winds, it is necessary to calculate the mean kinetic energy that results from the superposition of the wind velocity $v_w$ and the orbital velocity $v_o$ of a cluster star. In a reference frame at rest with respect to the central, massive object, the square of the velocity modulus of the material ejected from the orbiting star is
\begin{equation}
  W^2(\theta)=(v_w\cos \theta+v_o)^2+v_w^2\sin^2\theta\,,
  \label{vmod}
\end{equation}
where $\theta$ is the angle measured from the direction of the orbital motion. The mean squared velocity of the ejected material then is:
\begin{equation}
  \overline{v^2}=\frac{1}{4\pi} \int_0^\pi W^2(\theta)\,2\pi\sin\theta d\theta=v_w^2+v_o^2\,,
  \label{vmean}
\end{equation}
where $W^2(\theta)$ is given by equation (\ref{vmod}).

In order to proceed, we make the simplest possible assumption of circular orbits around the central black hole for the cluster stars. Then the orbital velocity $v_o$ is
\begin{equation}
  v_o=\sqrt{\frac{GM_{bh}}{R}}\,,
    \label{vo}
\end{equation}
where $G$ is the gravitational constant, and $R$ the spherical radius. Thus, the rate of wind kinetic energy injection (per unit time and volume) by the stellar winds is:
\begin{equation}
  {\dot e}_{kin}=\frac{n{\dot M}_w}{2}\left(v_w^2+\frac{GM_{bh}}{R}\right)\,,
  \label{ekin}
\end{equation}
where we have used equations (\ref{vmean}) and (\ref{vo}). The stellar winds also introduce gravitational potential energy into the combined cluster wind at a rate:
\begin{equation}
  {\dot e}_{pot}=-n{\dot M}_w \frac{GM_{bh}}{R}\,.
  \label{epot}
\end{equation}

The mass and energy input rates integrated in a volume out to a radius $R$ are equal to the integral over the $R$-surface of the mass and energy fluxes (which in our spherically symmetric models amounts to multiplying the fluxes by $4\pi R^2$). The mass flux is
\begin{equation}
  F_{mass}=\rho v\,,
  \label{fmass}
\end{equation}
where $\rho$ is the density and $v$ the velocity of the cluster wind. The energy flux is:
\begin{equation}
  F_{en}=\rho v\left(\frac{v^2}{2}+\frac{\gamma}{\gamma-1}\frac{P}{\rho}-\frac{GM_{bh}}{R}\right)\,,
  \label{fen}
\end{equation}
where $P$ is the gas pressure and $\gamma=c_p/c_v$ is the ratio of specific heats ($= 5/3$ for a monoatomic gas). The equations resulting from equating the volume integrals of the
mass (equation \ref{min}) and energy (equations \ref{ekin}-\ref{epot}) source terms with the surface integrals of the corresponding fluxes (equations \ref{fmass}-\ref{fen}) are given in the following section.

\subsection{The flow equations}
The resulting mass conservation equation is:
\begin{equation}
  4\pi R^2 \rho v=\frac{4\pi}{3}R^3 n {\dot M}_w\,,
  \label{m}
\end{equation}
and the energy conservation gives:
$$
   4\pi R^2 \rho v\left(\frac{v^2}{2}+\frac{\gamma}{\gamma-1}\frac{P}{\rho}-\frac{GM_{bh}}{R}\right)=
$$
\begin{equation}
  \frac{2\pi n \dot{M}_w}{3}\left(R^3v_w^2-\frac{3}{2}GM_{bh}R^2\right)\ \mbox{ ,}
  \label{en}
\end{equation}
where the right hand side of equation \ref{en} is the result of integrating
${\dot e}_{kin}$ + ${\dot e}_{pot}$ over the volume.

Also, we have to consider an equation of motion of the form:
\begin{equation}
  \rho v \frac{dv}{dR}=-\frac{dP}{dR}-n{\dot M}_wv-\rho\frac{GM_{bh}}{R^2}\,,
  \label{mot}
\end{equation}
where the three terms on the right hand side are the pressure gradient force, the drag force necessary for incorporating the stellar winds into the cluster wind flow and the gravitational attraction of the central black hole. Note that we have ignored the gravity force of the stars.

Equations (\ref{m}-\ref{mot}) are valid within the cluster radius, i.e., for $R\leq R_c$. The flow equations for $R>R_c$ are obtained by setting $R=R_c$ in the right-hand sides of equations (\ref{m}) and (\ref{en}), and $n=0$ in the right-hand-side of equation (\ref{mot}).

Combining equations (\ref{m}-\ref{en}) we obtain
\begin{equation}
  \rho=\frac{n{\dot M}_w R}{3v}\,,
  \label{mm}
\end{equation}
\begin{equation}
  c_s^2=\frac{\gamma-1}{2}\left(v_w^2-v^2+\frac{GM_{bh}}{2R}\right)\,,
  \label{cs}
\end{equation}
where $c_s=\sqrt{\gamma P/\rho}$ is the sound speed. Using these two relations, we can then turn the equation of motion (equation \ref{mot}) into a differential equation involving only the velocity $v$ of the gas:
\begin{equation}
  \frac{dv}{dR}=\frac{2v}{R}\frac{(1+5\gamma)v^2+(\gamma-1)v_w^2+2\gamma GM_{bh}/R}
       {2(\gamma-1)v_w^2-2(\gamma+1)v^2+(\gamma-1)GM_{bh}/R}\,.
       \label{vr}
\end{equation}
An integration of equation (\ref{vr}) gives the velocity of the cluster wind as a function of radius $R$, and substituting this solution into equations (\ref{mm}) and (\ref{cs}) we obtain the spatial dependence of the density and sound speed (respectively) of the wind in the $R\leq R_c$ region.

Outside the edge of the cluster (for $R>R_c$, see
the text following equation \ref{mot}), the density, sound speed and velocity of the cluster wind are given by:
\begin{equation}
  \rho=\frac{n{\dot M}_w R_c^3}{3R^2v}\,,
  \label{mmc}
\end{equation}
\begin{equation}
  c_s^2=\frac{\gamma-1}{2}\left[v_w^2-v^2+GM_{bh}\left(\frac{2}{R}-\frac{3}{2R_c}\right)\right]\,,
  \label{ccs}
\end{equation}
where $c_s=\sqrt{\gamma P/\rho}$ is the sound speed. Using these two relations (and setting $n=0$ in equation \ref{mot}), we can then turn the equation of motion into a differential equation involving only the velocity $v$ of the gas:
$$\frac{dv}{dR}=\frac{2v}{R}\times$$
\begin{equation}
\frac{(\gamma-1)(v_w^2-v^2)+GM_{bh}\left[(2\gamma-3)/R-3(\gamma-1)/(2R_c)\right]}
       {(\gamma+1)v^2-(\gamma-1)v_w^2+(\gamma-1)GM_{bh}[3/(2R_c)-2/R]}\,.
       \label{vrc}
\end{equation}

\subsection{The dimensionless flow equations}
We use the radius at which the circular orbital velocity is equal to the velocity of the stellar winds
\begin{equation}
  R_0=\frac{GM_{bh}}{v_w^2}\,,
\label{r0}
\end{equation}
and $v_w$ to define a dimensionless radius $r=R/R_0$, and velocity $u=v/v_w$ . The dimensionless
radius of the cluster is then $r_c=R_c/R_0$. In terms of these dimensionless variables, the flow equations for $r\leq r_c$ (see equations \ref{mm}-\ref{vrc}) become:
\begin{equation}
  \frac{\rho}{\rho_0}=\frac{r}{u}\,,
  \label{amm}
\end{equation}
\begin{equation}
  \left(\frac{c_s}{v_w}\right)^2=\frac{\gamma-1}{2}\left(1-u^2+\frac{1}{2r}\right)\,,
  \label{acs}
\end{equation}
\begin{equation}
  \frac{du}{dr}=\frac{2u}{r}\frac{(1+5\gamma)u^2+\gamma-1+2\gamma /r}
       {2(\gamma-1)-2(\gamma+1)u^2+(\gamma-1)/r}\,,
       \label{avr}
\end{equation}
where $\rho_0=n{\dot M}_wGM_{bh}/(3v_w^3)$ in equation (\ref{amm}).

For $r>r_c$ (outside the cluster radius see equations \ref{mmc}-\ref{vrc}), the dimensionless flow equations can be written as:
\begin{equation}
  \frac{\rho}{\rho_0}=\frac{r_c^3}{r^2u}\,,
  \label{ammc}
\end{equation}
\begin{equation}
  \left(\frac{c_s}{v_w}\right)^2=\frac{\gamma-1}{2}\left(1-u^2+\frac{2}{r}-\frac{3}{2r_c}\right)\,,
  \label{accs}
\end{equation}
\begin{equation}
  \frac{du}{dr}=\frac{2u}{r}
  \frac{(\gamma-1)(1-u^2)+(2\gamma-3)/r-3(\gamma-1)/(2r_c)}
       {(\gamma+1)u^2-(\gamma-1)+(\gamma-1)[3/(2r_c)-2/r]}\,.
  \label{avrc}
  \end{equation}

\begin{figure}
\centering
\includegraphics[width=\columnwidth]{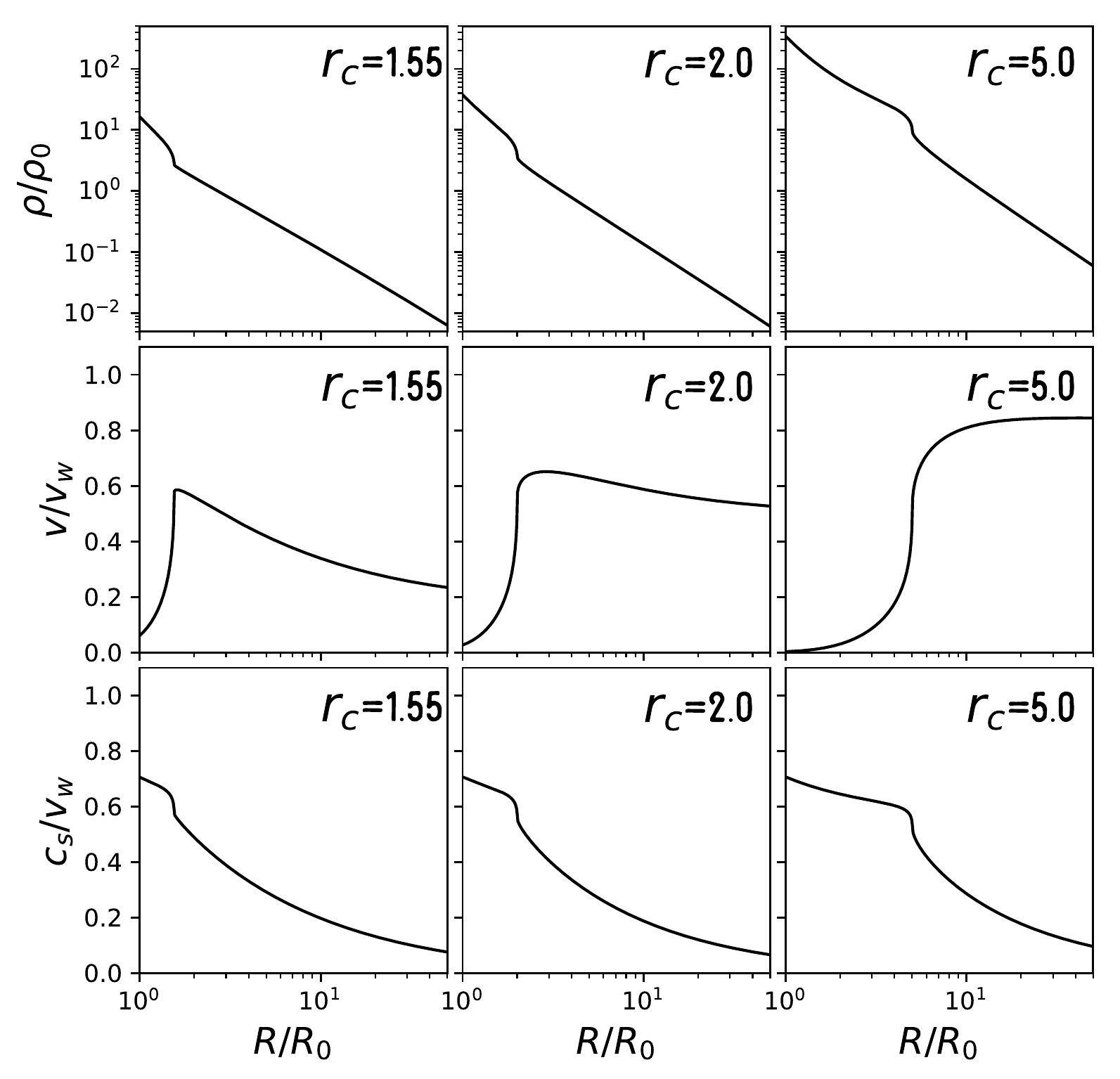}
\caption{Radial profiles of density (top row), velocity (middle row, and sound speed (bottom row). The solutions were obtained for different cluster radii: $r_c=1.55$ (left column), $r_c=2.0$ (middle column), and $r_c =5.0$ (right column).}
\label{f1}
\end{figure}
%

\begin{figure*}
    \centering
    {{\includegraphics[width=0.4\textwidth]{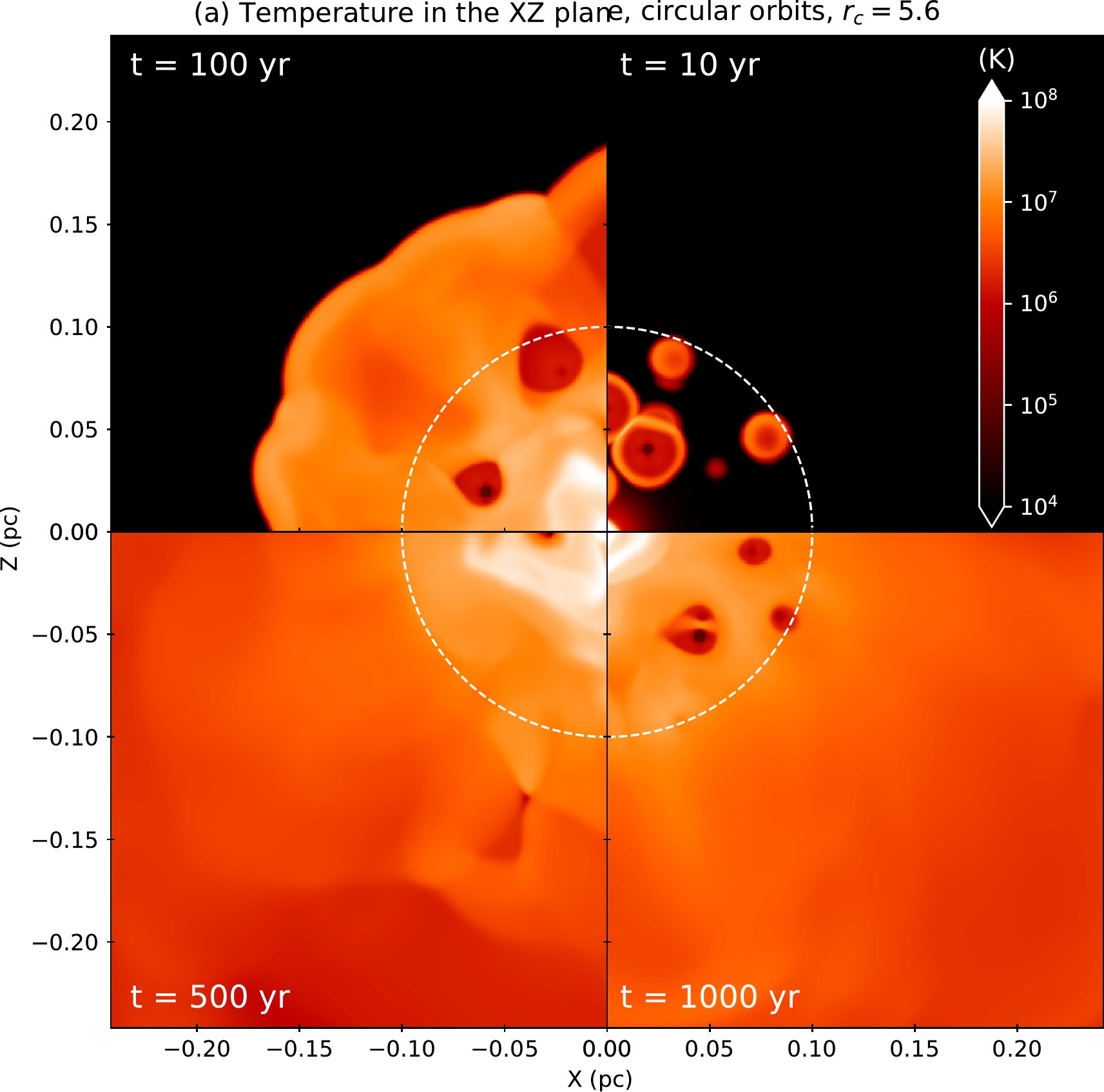}}} 
    {{\includegraphics[width=0.4\textwidth]{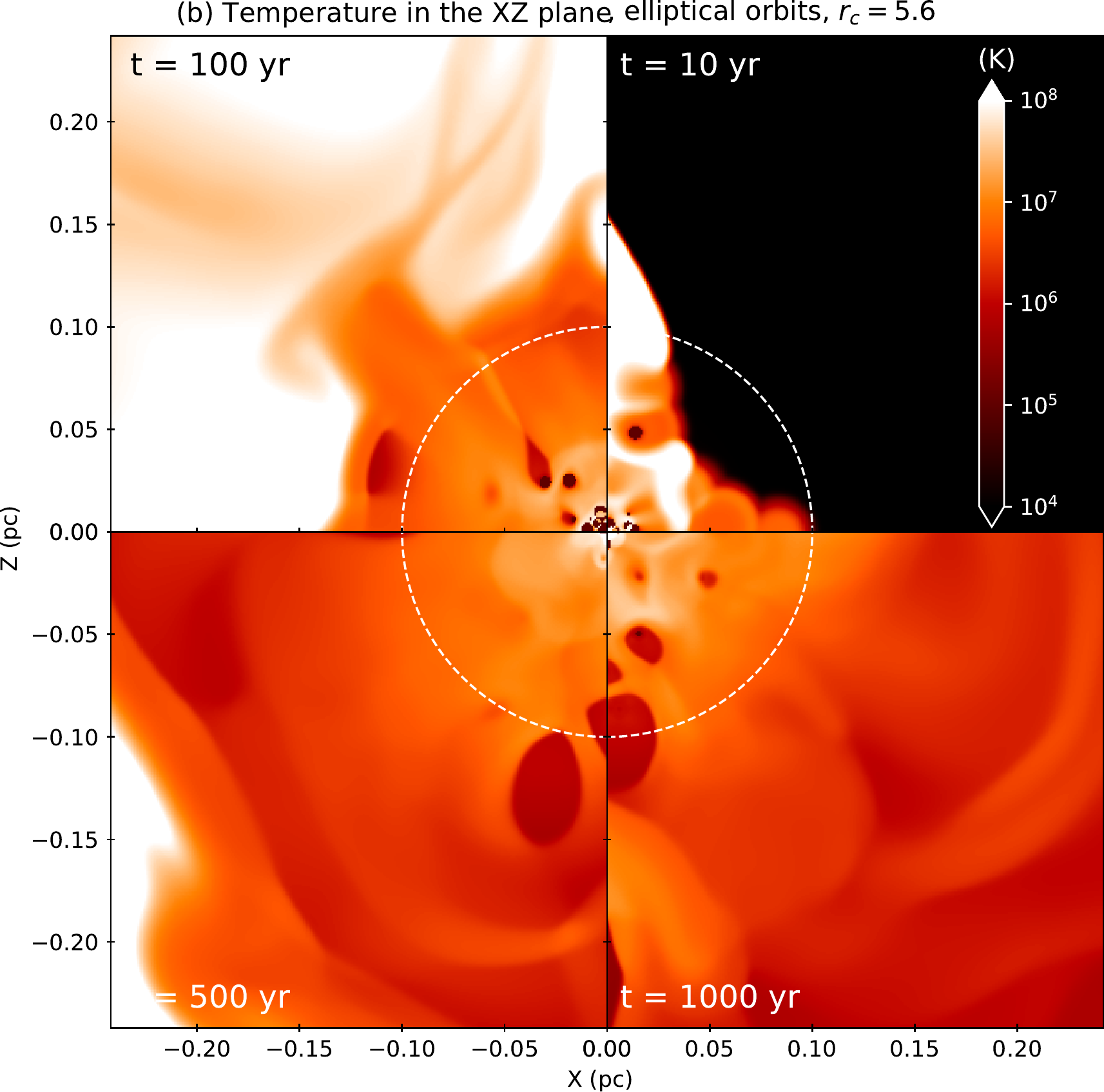}}}
    {{\includegraphics[width=0.4\textwidth]{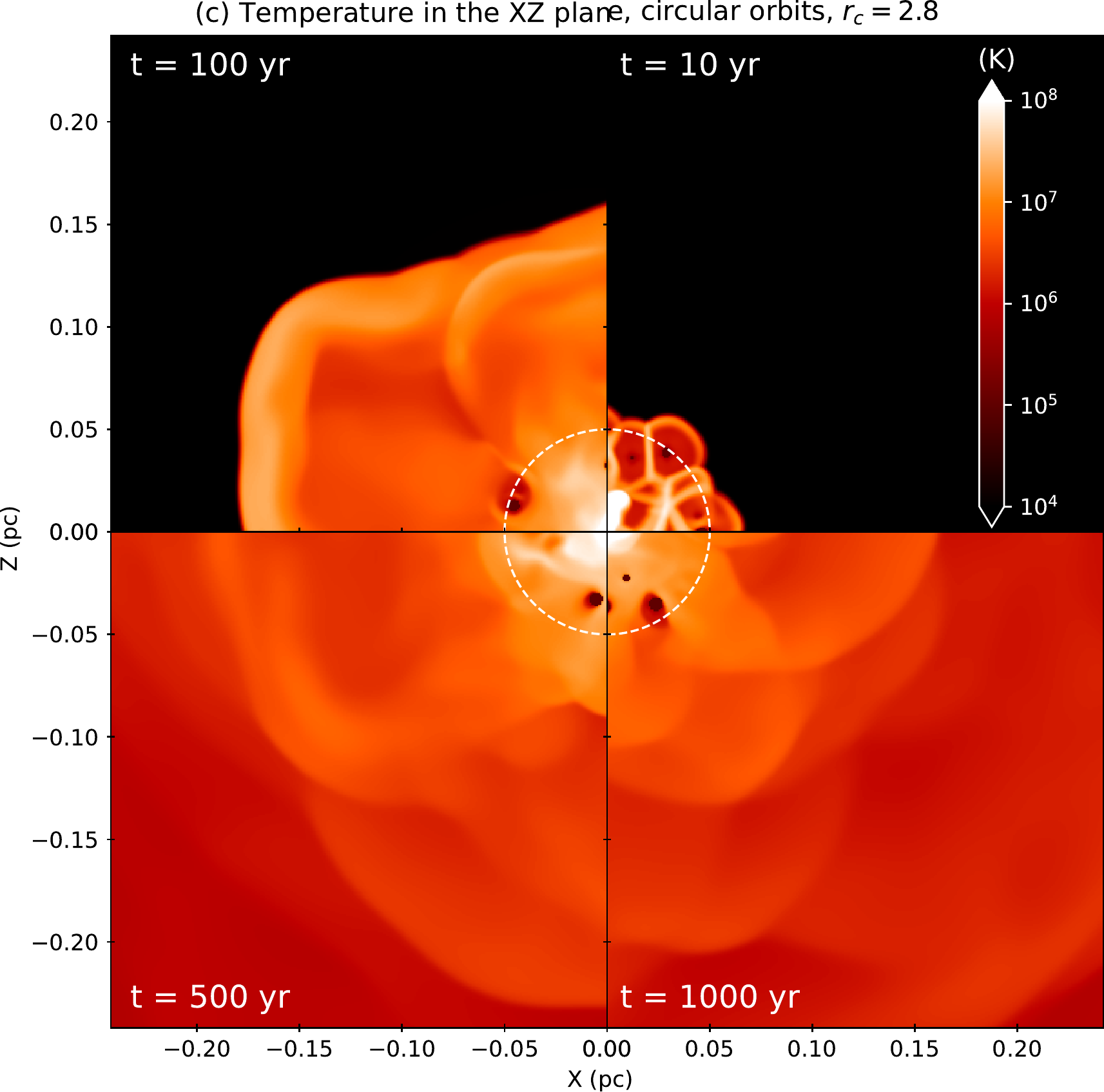}}}
    \caption{We show the evolution of the temperature in the Y-midplane for four integration times in the simulation: t=10 (upper-right panels), 100 (upper-left panels), 500 (lower-left panels) and 1000~yr (lower-right panels).
    The white circle shows the size of the stellar cluster with radius $R_c$. The units of the color bars are given in K. The Figure labeled ``(a)'', shows the temperature evolution for the circular orbit case and $r_c=5.6$, the Figure labeled ``(b)'', shows the evolution of the temperature for the elliptic orbit case and $r_c=5.6$, and the Figure labeled ``(c)'' shows the evolution of the circular orbit case and $r_c=2.8$.}
    \label{fig:temp}
\end{figure*}

\subsection{Analytical considerations}
An inspection of equation (\ref{avr}) shows that for a cluster with a finite velocity close to the origin (e.g. near the location of the massive object at the centre of the cluster) the $2\gamma/r$ and $(\gamma-1)/r$ dominates over the other terms in the numerator and denominator (respectively). Neglecting these other terms, for $r\ll 1$ we then have:
\begin{equation}
  \frac{du}{dr}=\frac{4\gamma u}{(\gamma-1)r}\,,
  \label{duc}
\end{equation}
which has the solution
\begin{equation}
  u=\left(Ar\right)^{4\gamma/(\gamma-1)}\,,
  \label{us}
\end{equation}
where $A$ is an integration constant. We should note that for $\gamma=5/3$,
the exponent in equation (\ref{us}) has a value of 10.

If for larger $r$ the velocity $u$ continues to increase, $du/dr$ will eventually diverge. This divergence occurs at the point in which the denominator of equation (\ref{avr}) is equal to zero. From this condition, we obtain the relation:
\begin{equation}
  u_d^{2}=\frac{\gamma-1}{\gamma+1}\left(1+\frac{1}{2r_d}\right)\,,
  \label{ud}
\end{equation}
where $u_d$ is the (dimensionless) cluster wind velocity at the radius $r_d$ at which
$du/dr$ diverges. Substituting equation (\ref{ud}) into equation (\ref{acs}), we see that $c_s/v_w=u_d$ at $r=r_d$. Therefore, at the point $r_d$ in which $du/dr$ diverges, the flow is sonic.

Now, if we look at the $r>r_c$ solution (equations \ref{ammc}-\ref{avrc}), we see that for $r=r_c$ the condition for a divergence of $du/dr$ gives
\begin{equation}
  u_c^{2}=\frac{\gamma-1}{\gamma+1}\left(1+\frac{1}{2r_c}\right)\,.
  \label{ucc}
\end{equation}
Therefore, if for the inner ($r\leq r_c$) solution we choose a radius $r_d=r_c$ (for $du/dr$ divergence), we the have a continuous transition to the outer ($r>r_c$) solution. This matching between the inner and outer solutions with a sonic point at $r=r_c$ is equivalent to the one of the ``classical'' cluster wind solution (i.e., with no gravity), see, e.g. \citet{can00}.

We should point out that in principle, the inner solution
  could end at a cluster radius $r_c<r_d$, so that $du/dr$ does not
  diverge within the cluster. However, this would produce a subsonic
  velocity at $r_c$ (see equation \ref{ud} and the text following this
  equation). The $r>r_c$ solution (see equation \ref{avrc}) would then
  have a subsonic initial condition (at $r=r_c$). However, it is a well known
  result that an adiabatic, spherical flow does not have a sonic
  transition (see, e.g., the book of Lamers \& Casinelli 1999), and the
  flow will never reach supersonic velocities.

The fact that the fluid has an effective $\gamma < 5/3$
in the classical Spitzer isothermal wind (for a single star) is due
to the very efficient thermal conduction that is able to maintain a close to isothermal flow. For the physical conditions of the interaction of several winds, the thermal conduction is not as efficient, and thus an adiabatic treatment is more adequate.

Therefore, the only
  possibility left is to have the critical point of the full,
  inner+outer solution coinciding with the cluster radius. This
  result does not hold for an isothermal flow, or for a wind with
  thermal conduction, in which one can in principle have the
  sonic point within the outer (and possibly also within the inner)
  solution.

Another interesting point that can be made through an inspection of the flow equations is as follows. For $r\to \infty$ the wind will reach a constant, terminal velocity. Therefore, $du/dr\to 0$ for $r\to \infty$. Looking at the numerator of equation (\ref{avrc}) we see that $du/dr\to 0$ implies that we either have $u\to 0$ (this would be a ``stalled wind'' solution) or that
\begin{equation}
  u^{2}\to u_\infty^{2}=1-\frac{3}{2r_c}\,.
  \label{uinf}
\end{equation}
Therefore, for $r_c>3/2$ we will have the terminal wind velocity given by equation (\ref{uinf}). Using equations (\ref{vo}) and (\ref{r0}, and
noting that $r_c=R_c/R_0$, this condition can be written
as $v_w>\sqrt{3/2}v_o(R_c)$, where $v_o(R_c)$ is the circular orbital
velocity at the outer boundary of the cluster.
As will be discussed in the following section, there are no full
cluster wind solutions when this condition is not satisfied.

\subsection{Flow solutions}
The inner flow ($r\leq r_c$, within the cluster) can be straightforwardly obtained by numerically integrating equation (\ref{avr}), starting with an off-center $(u,r)$ initial condition
found through the $r\ll 1$ flow solution of equation (\ref{us}) with an appropriately chosen value for the $A$ integration constant (this choice of $A$ is done by trying different values until the desired cluster radius $r_c$ is obtained). The numerical integration is continued until the flow velocity $u(r)$ becomes less or equal than the velocity expected for the cluster edge (the velocity $u_c$
of equation \ref{ucc}, computed at $r_c=r$).

When this condition is met, we have arrived at the
edge of the cluster, and therefore switch to an integration of the outer flow equation (\ref{avrc}), starting at the current $(u,r)=(u_c,r_c)$ values.

Alternatively, one can use the analytic integral:
\begin{equation}
  f(r,u)=u^2+\frac{2 u_c^{\gamma+1}}{(\gamma-1)u^{\gamma-1}}\left(\frac{r_c}{r}\right)^{2(\gamma-1)}-\frac{2}{r}+\frac{2}{r_c}-
  \frac{\gamma+1}{\gamma-1}u_c^2=0\,.
  \label{ff0}
\end{equation}

This solution can be straightforwardly obtained by noting that the outer flow follows an adiabat (i.e., that $c_s^{2}\propto \rho^{\gamma-1}$), and combining this condition with equations (\ref{m}) and (\ref{en}). In order to obtain $u(r)$, we fix values for $r$ (starting at $r=r_c$, where the outer flow starts) and numerically find the values of $u$ for which $f(r,u)=0$.

Interestingly, for $r_c\leq 3/2$ equation (\ref{ff0}) has no zeroes for $r>r_c$. Therefore, in order to have a steady cluster wind the $r_c>3/2$ condition has to be met (see also the discussion following
equation \ref{uinf}). For $r_c>3/2$, for all radii $r>r_c$ there are two values of $u$ for which equation (\ref{ff0}) has zeroes: one of them supersonic (the wind solution) and the other wind subsonic (a ``stalled wind'' solution). The supersonic $u(r)$ solution coincides with the numerical integration of the outer wind differential equation
(equation \ref{avrc}) described above.

In Figure 1 we show the radial structure of the flow velocity, sound speed and density obtained for clusters with $r_c=1.55$, 2 and 5 (for a flow with $\gamma=5/3$). For the clusters with $r_c=1.55$ and 2 we find
that the flow velocity initially grows with $r$, reaches a peak (outside the cluster radius) and then decreases monotonically to attain the asymptotic value given by equation (\ref{uinf}). The $r_c=5$
cluster has a monotonically increasing $u(r)$ (also converging at large radius to the appropriate asymptotic value).

The sound speed has a divergence at the origin, followed by a sharp decrease which produces a peak extending to $r\sim 1$ (a radius $R\sim R_0$, see equation \ref{r0}). The sound speed then has a plateau (with a sound speed $c_s\sim 0.6 v_w$) extending out to the cluster radius $R_c$, and a monotonic decrease with $R$ outside the cluster. The density is a monotonically decreasing function of radius, with a sharper decrease close to the cluster radius (see Figure 1).

We should point out an important difference between our cluster
wind solutions and the ones of \cite{silich:08}. These authors
find that in order to obtain an outflowing cluster wind, they need
to have an inner region in which the flow is directed inwards,
accreting onto the SMBH. \cite{qua:04} found the same result
through an appropriate integration of the time-dependent
equations.

Our model differs from the
ones of \cite{silich:08} and \cite{qua:04} because it
includes the effect of the orbital motion of the stars in
the stellar wind energy equation, see our equations 3-5.
Because of this difference, we do
obtain an outwards directed wind at all radii, provided that
the condition $r_c=R_c/R_0> 3/2$ is met (see equation 29).


Our equations do not appropriate model the region close to the central BH, i. e. we do not include accretion into the SMBH. However, in this inner region, an infall into the central object is unavoidable.

\section{Numerical simulations setup}
\subsection{The code: {\sc guacho} + N-body}
\label{sec:code}
We carried out 3D-grid hydrodynamic simulations with the {\sc guacho} code
\citep{esquivel:09,esquivel:13}. The code solves the ideal hydrodynamic equations in a uniform Cartesian grid with a second-order Godunov method with an approximate Riemann solver (in this case we use the HLLC solver, \citealt{toro:99}), and a linear reconstruction of the primitive variables using a minmod slope limiter to ensure stability.
We assume a gas of pure hydrogen, along with the gasdynamic equations we solve a rate equation for neutral hydrogen of the form
\begin{equation}
    \frac{\partial n_{HI}}{\partial t} + \nabla (n_{HI}\mathbf{u}) =
    n_{e} n_{HII} \alpha(t) - n_{HI} n_{HII} c(t)
    \mbox{ ,}
\end{equation}
where $n_{HI}$ is the neutral hydrogen density, $n_{HII}$ is the ionized hydrogen density, and $n_{e}$ is the electron density. In this equation we denote \textbf{u} as the flow velocity, $\alpha(t)$ is the recombination (case B) coefficient, and $c(t)$ is the collisional ionization coefficient.
The energy equation and the hydrogen continuity equation are integrated forward in time, without the source terms in the hydrodynamic timestep. Instead, the source terms are added in a semi-implicit timestep with which the ionization fraction is updated.
We do not include cooling since the cooling times at the temperatures we are working with, are much larger than the evolution time of the simulation.

In order to include the effects of the stellar orbits in the calculation we coupled an N-Body module to the {\sc guacho} code. The N-body module is a version of the {\sc Varone} code \citep{lora:09}. Since we are dealing with a small number of particles/stars ($N=100$), we are able to compute all the gravitational interactions between them, and thus we use a direct variant of the {\sc Varone} code/module.
The N-body module updates the positions of the wind sources at every timestep of the hydrodynamic code, which also adds the orbital velocity to the wind velocity.

 \begin{figure}
 \centering
 \includegraphics[width=8.5cm]{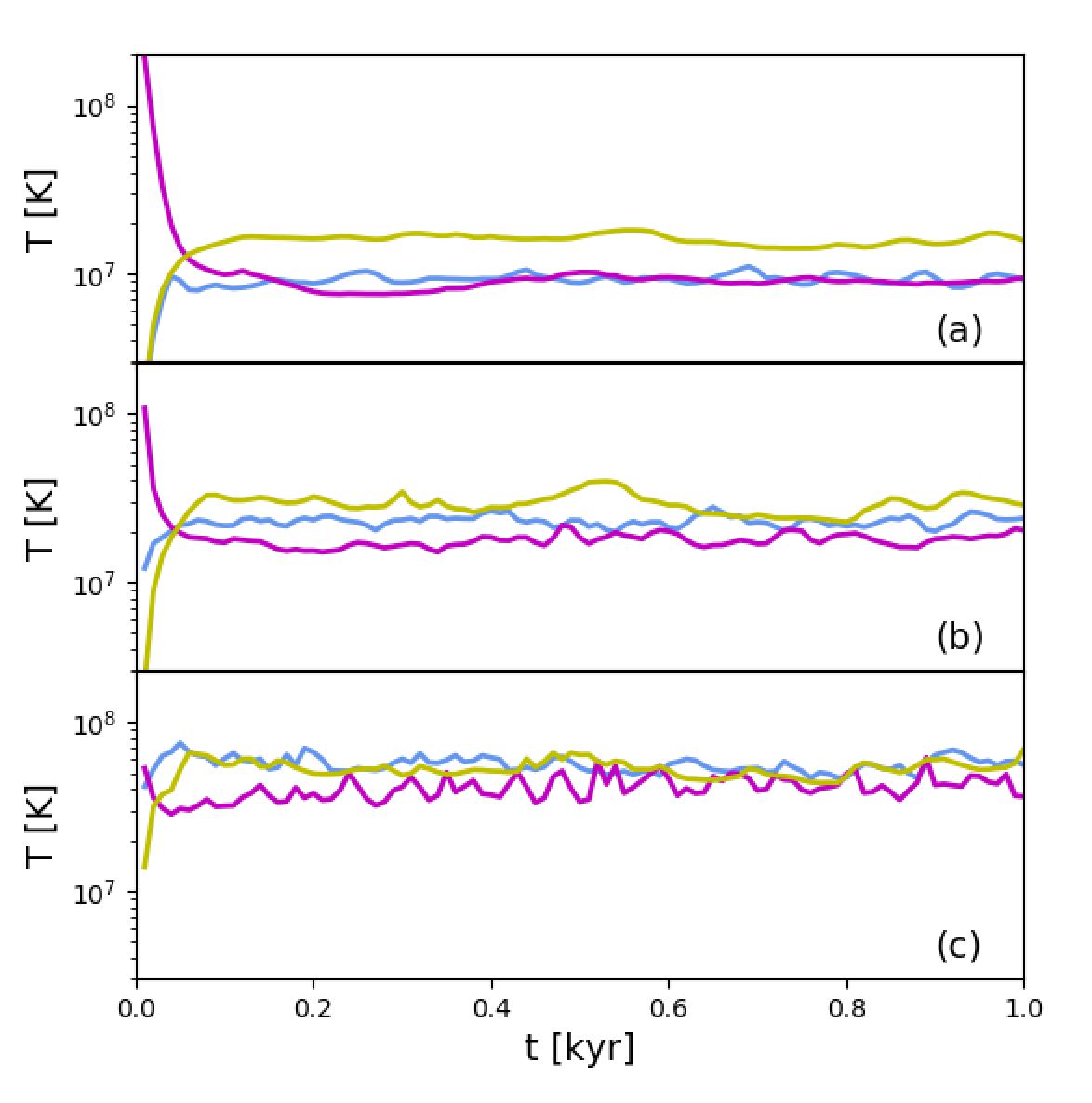}
 \caption{In this Figure we show the average of the
 temperature inside a sphere of a different radii.
 The blue lines correspond to the circular model, the magenta lines correspond to the elliptical model, and the yellow lines correspond to the small circular model.
 In panel $a$ we show the average temperature inside a sphere of
 radius $R\simeq 0.1$ pc (which is equivalent to $r_c$ in the circular and elliptical models).
 In panel $b$ we show the average temperature inside a sphere of
 radius $R\simeq 0.05$ pc (which is equivalent to $r_c/2$ for the circular and elliptical models).
  In panel $c$ we show the average temperature inside a sphere of
 radius $R\simeq 0.025$ pc (which is equivalent to $r_c/5$ for the circular and elliptical models).
}
\label{fig:temp_vs_time}
\end{figure}

\begin{figure*}
    \centering
    {{\includegraphics[width=0.4\textwidth]{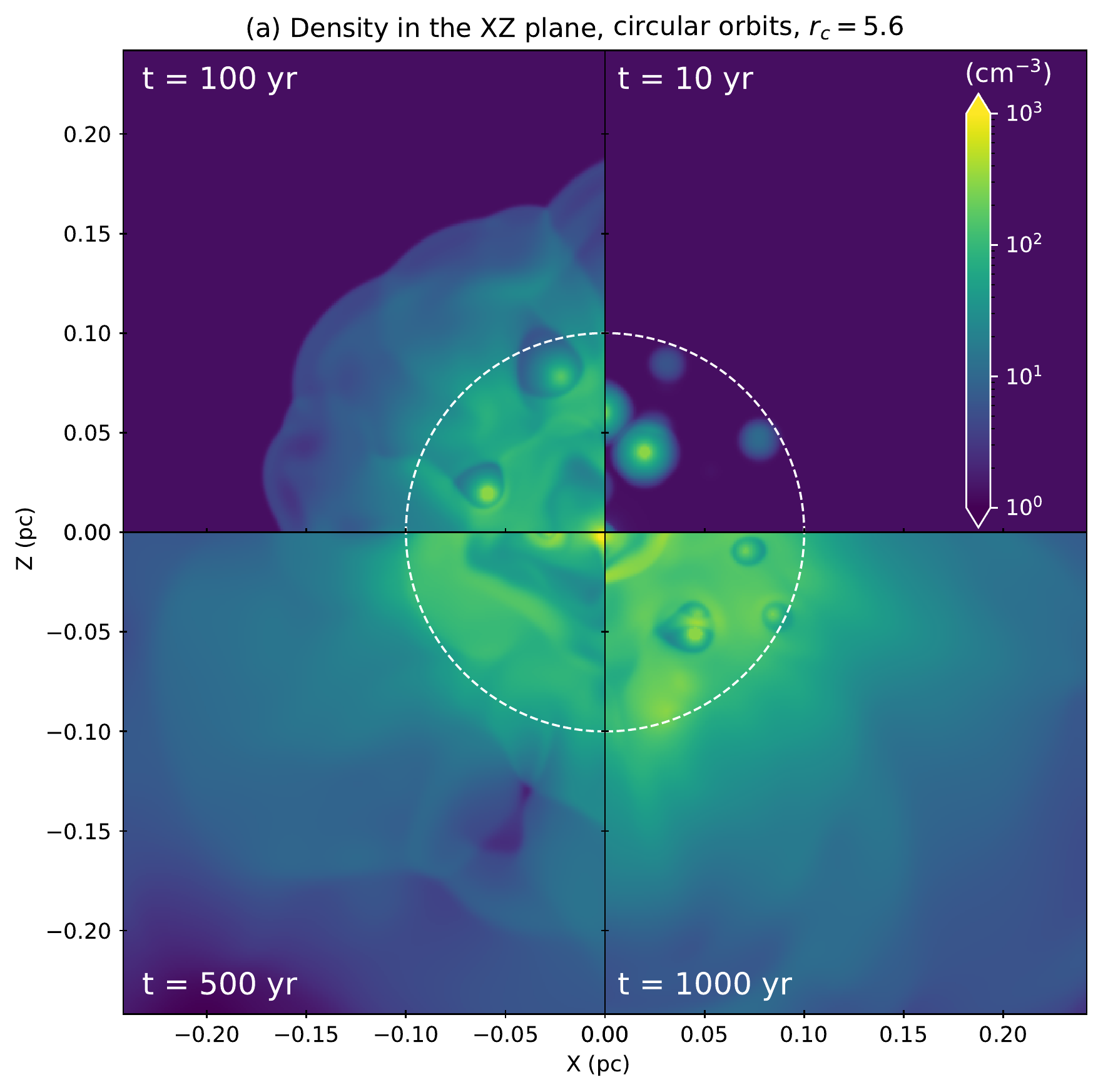} }}  
    {{\includegraphics[width=0.4\textwidth]{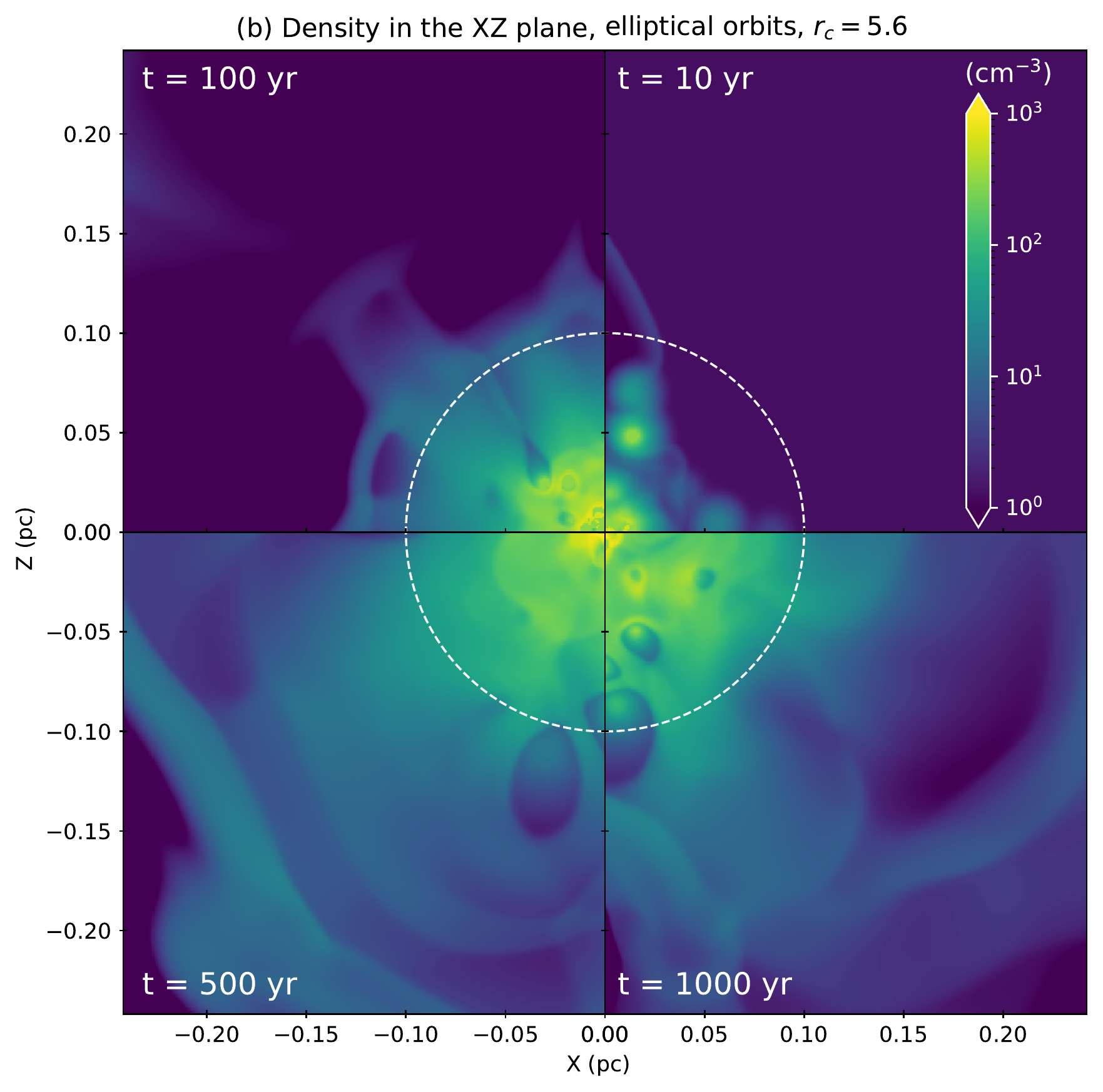} }}
    {{\includegraphics[width=0.4\textwidth]{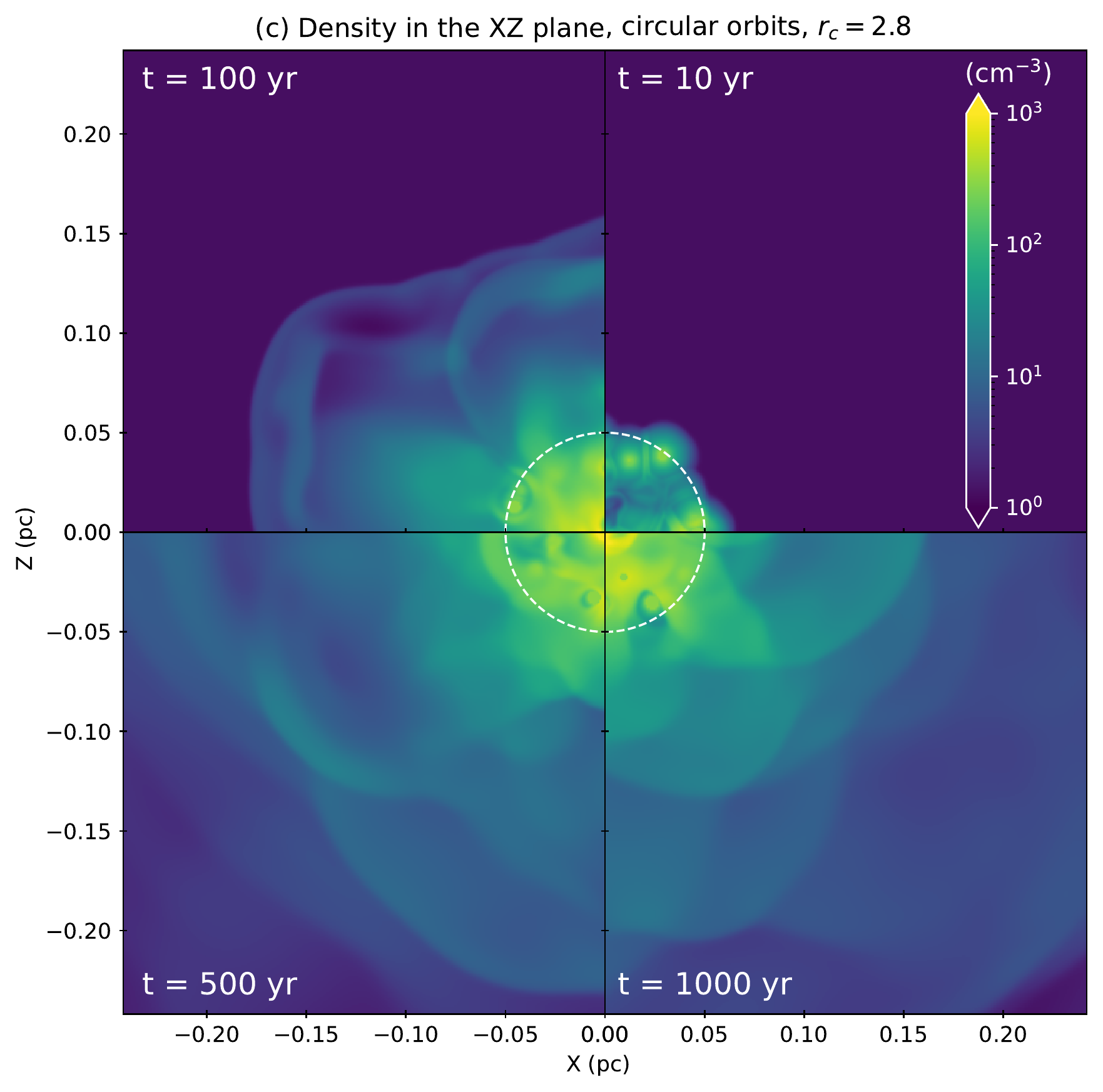} }}
    \caption{We show the evolution of the density in the Y-midplane for four integration times in the simulation;  t=10 (upper-right panels), 100 (upper-left panels), 500 (lower-left panels) and 1000~yr (lower-right panels).
    The white circle shows the size of the stellar cluster with radius $R_{c}$. The units of
    the color bar are given in particles cm$^{-3}$. The Figure labeled ``(a)'', shows the density evolution for the circular orbit case and $r_c=5.6$, the Figure labeled ``(b)'' shows the evolution of the density for the elliptic orbit case and $r_c=5.6$, and the Figure labeled ``(c)'' shows the evolution for the circular orbit case and $r_c=2.8$.}
    \label{fig:density}
\end{figure*}

\subsection{Cluster+SMBH initial conditions} %
\label{sec:IC}                               %
We model a star cluster containing $100$ stars. The computational domain has a physical size of $1.5\times10^{18}\times 1.5\times 10^{18}\times 1.5\times 10^{18}$~cm ($0.5\times0.5\times0.5$ pc) along the $x$-,$y$-, and $z$-axes, which is resolved with a uniform Cartesian grid of 600$\times$600$\times$600 cells.

We impose an isotropic wind for each one of these stars. The stellar winds are imposed to be fully ionized, with a mass loss rate of $\dot{M}=1\times10^{-6}$~M$_{\odot}$/yr.
The temperature associated to the star-winds is $T_{wind}=1\times10^5$~K, and the velocity of the stellar wind is $v_{wind}=1\times10^8$~cm/s. The star-wind is centered at each star position within a radius $r_{wind}=1\times10^{16}$~cm. The outer boundary condition on the surface of these spheres is an outward flowing wind. In this work we do not impose an accreting flow into the black hole. The region where the wind sources are injected is significantly larger than the stellar radius. Therefore, in their orbital motion wind sources occasionally overlap, when this occurs we superimpose the winds of the overlapping sources.

We add a massive particle in the cluster's center, with a mass
$M_{BH}=4\times10^{6}$~M$_{\odot}$ mimicking a SMBH with a mass similar to the one of the SMBH in the center of the Galaxy.

We generate the initial positions and velocities of the $N$-stars considering the mass of the SMBH in the center of the star cluster.

The initial position and orientation of the stars orbits are set randomly within a radius $R_c$.
For the initial position we draw three random numbers ($x,~y,~z$) from a uniform distribution between $-R_c$ and $R_c$, if the position with respect to the center  ($r_{*}=\sqrt{x^2+y^2+z^2}$) is larger than $R_c$ we discard the random numbers and repeat the procedure until the $N$ stars are placed. We constructed three distributions, two with $R_c = 3\times 10^{17}~\mathrm{cm}$ (corresponding to a dimensionless $r_c = R_c/R_0=5.6$) and one with $R_c = 1.5\times 10^{17}~\mathrm{cm}$ ($r_c=2.8$).

For the initial velocities of the stars, once the positions of each of the stars are computed, we calculated the circular velocity of each star considering only the mass of the SMBH ($v_{c,*} = (GM_{BH}/r_{*})^{1/2}$). Then, we chose a random number between zero and 1, and multiply it by the circular velocity at $r_{*}$, for each of the stars. As a result we have random eccentricities of the orbits of the stars.

For the orbital velocities, we built three different cases: in the first two cases we impose a circular orbit to each of the stars in the cluster. To do this we compute the magnitude of the circular velocity at the initial position of each star and we add this velocity, projected onto a random orientation in the plane that is perpendicular to the radial position vector. We ran two models with circular orbits, one with $r_c=5.6$ and the second one with  $r_c=2.8$.
In the third model we impose eccentric orbits for the stars in the cluster for a distribution with $r_c=5.6$. The mass of each of the stars was set to $M_{*}=1$~M$_{\odot}$ for both circular and eccentric orbit cases.
The environment is initially at rest, and consists of neutral hydrogen, with a density $\rho_{env}=2.16\times10^{-24}$~g~cm$^{-3}$ and a temperature $T_{env}=1\times10^4$~K.

\section{Results}
\label{sec:results}
We allowed the models to run from the initial conditions described in
\ref{sec:IC} to an evolutionary time $t=1$~kyr, for our initial circular and eccentric orbit cases.

In Figure~\ref{fig:temp} we show the time evolution of the Y-midplane temperature for four integration times: $10$, $100$, $500$ and $1000$~yr. In (a) we show the case where the orbits are circular with $r_c=5.6$, in (b) the case where the orbits are eccentric with $r_c=5.6$, and in (c) the case with circular orbits and $r_c=2.8$. We show the radius of the cluster, $R_{c}$, as a white circle. We observe that at an integration time $t=10$~yr, almost all of the gas inside the cluster has raised its temperature to at least $\sim10^{7}$~K. In the eccentric case (Figure~\ref{fig:temp}b) the temperature is somewhat higher.

In order to study quantitatively how the temperature increases inside the star cluster radius of the circular and elliptical case ($R_c=3\times10^{17}$ cm) as time evolves, we computed the average temperature taking into account each computational cell inside this radius, and repeat the procedure for all the snapshots in our simulation.
In Figure \ref{fig:temp_vs_time} we plot the temperature as a function of the integration time within a radius $R\simeq 0.1$ pc (panel $a$), $R\simeq 0.05$ pc (panel $b$), and $R\simeq 0.025$ pc (panel $c$). The color code in  \ref{fig:temp_vs_time} is the same in the three panels: the color blue represents the circular case (with $R_c=3\times10^{17}$ cm), the magenta color represents the elliptical case (with $R_c=3\times10^{17}$ cm), and the
yellow color represents the small circular case (with $R_c=1.5\times10^{17}$ cm).

\begin{figure*}
    \centering
    {{\includegraphics[width=0.4\textwidth]{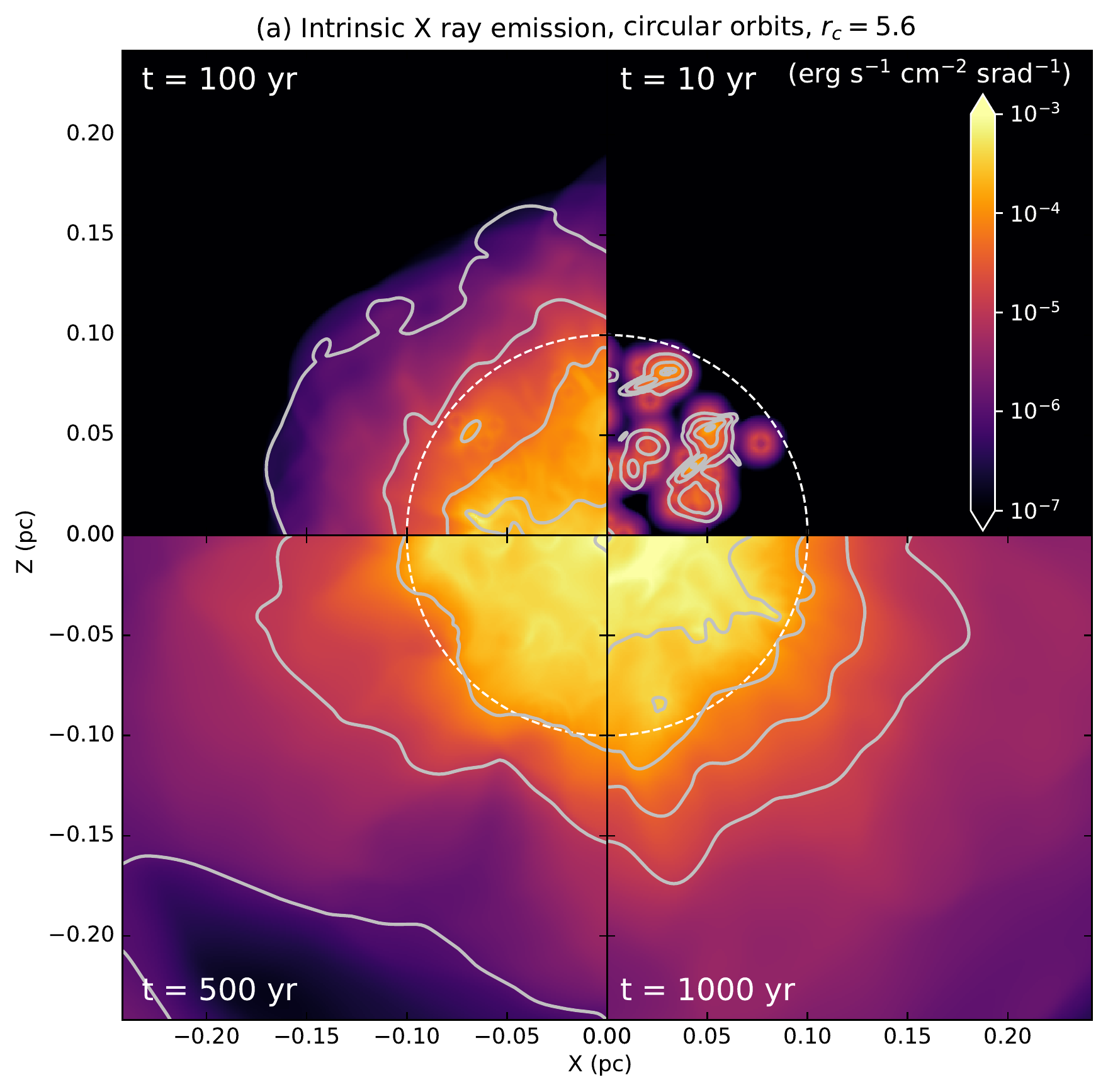} }}  
    {{\includegraphics[width=0.4\textwidth]{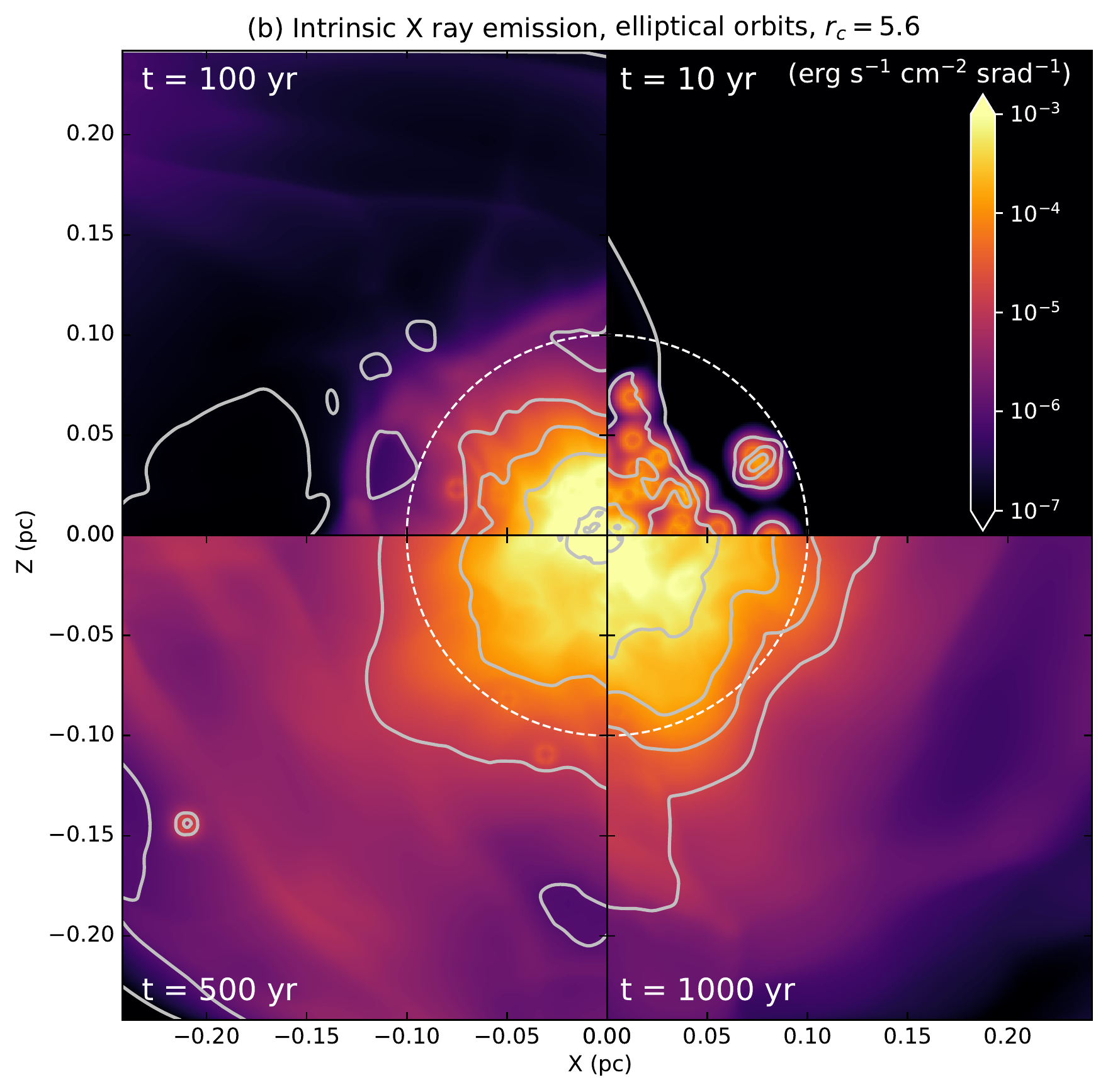} }}
    {{\includegraphics[width=0.4\textwidth]{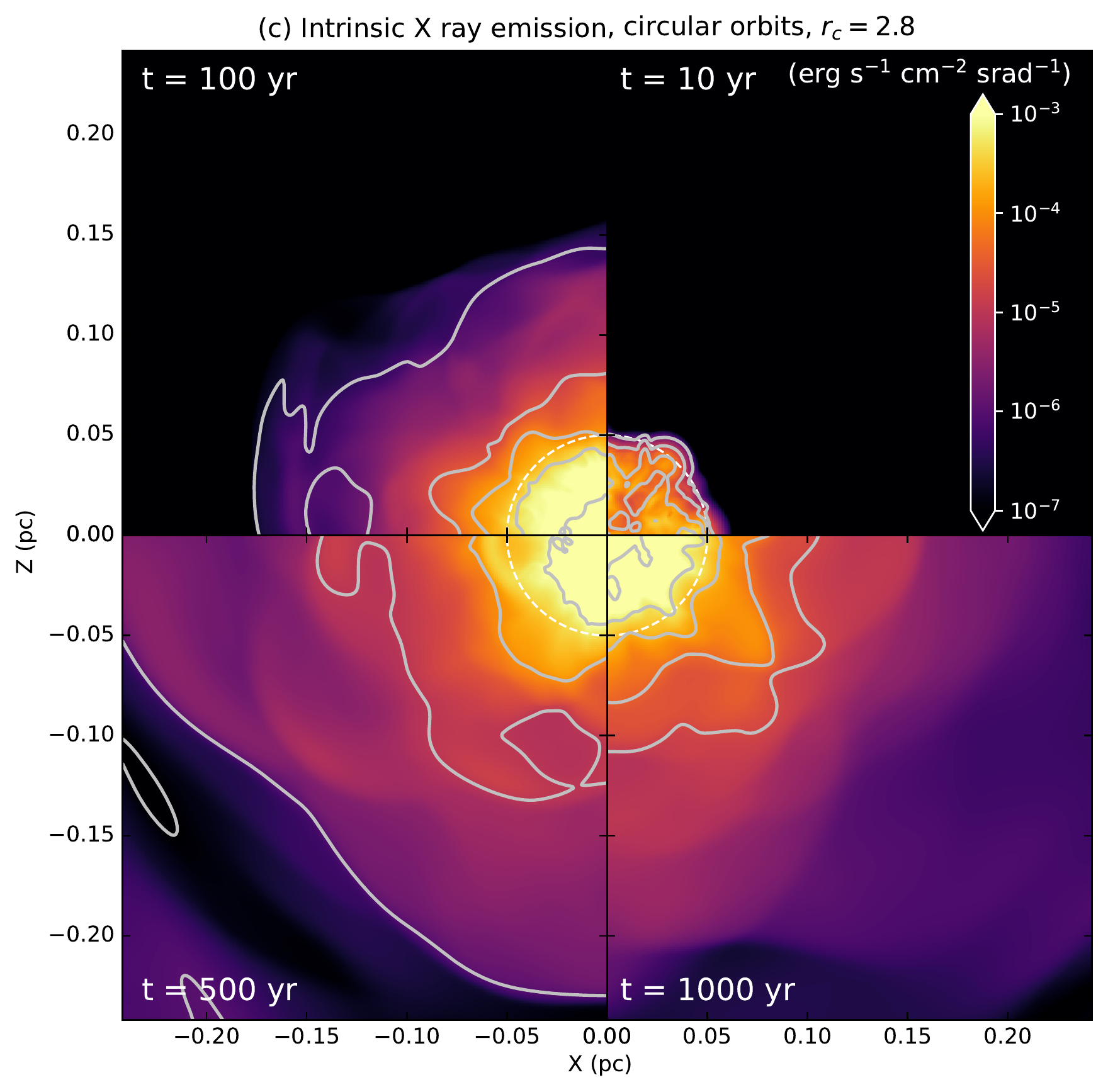} }}

     \caption{We show the evolution of the X-ray emission in the
     Y-plane for four integration times in the simulation;
     t=10 (upper-right panels), 100 (upper-left panels), 500 (lower-left panels) and 1000~yr (lower-right panels). The color palette shows
     the soft X-ray emission at $0.2$ to $2$ keV. The six white contours show the
     hard X-ray emission at $2-10$ keV, logarithmically spaced from $10^{-7}$
     to $10^{-3}$ erg s$^{-1}$ cm$^{-2}$ srad$^{-1}$.}
    \label{fig:xray}
\end{figure*}

In the circular orbit case, the average temperature value over all evolution times, inside $R\simeq0.1$ pc is $0.91\times10^7$~K. The average temperature value inside a radius $R\simeq0.05$ pc is $2.28\times10^7$~K, and
the temperature in the inner part of the cluster (inside a radius $R=\simeq0.025$ pc) averaged over all times is $5.62\times10^7$~K.

In the elliptic case, the average of the temperature over all times, inside $R\simeq0.1$ pc is $1.18\times10^7$~K. The temperature averaged over all times inside $R\simeq0.05$ pc is $1.89\times10^7$~K, and the temperature in the inner part of the cluster (inside a radius $R\simeq0.025$ pc) is $4.09\times10^7$~K.

In the small circular orbit case, the average of the temperature over all times, inside $R\simeq0.1$ pc is $1.55\times10^7$~K. The temperature averaged over all times inside $R\simeq0.05$ pc is $2.85\times10^7$~K, and the temperature in the inner part of the cluster
(inside a radius $R\simeq0.025$ pc) is $5.28\times10^7$~K.

In order to compare the temperature averaged inside $R\simeq0.025$ pc (panel $c$ in Figure \ref{fig:temp_vs_time}) for the circular and elliptical models, we define the ratio $\aries_{e,c}$ as
\begin{equation}
    \aries_{e,c} = \left| 1 - \frac{T_{e,c}}{T_{c,c}} \right| = 0.27  \mbox{ .}
    \label{ec:aries}
\end{equation}
Where $T_{e,c}$ is the average temperature in the elliptical model, and $T_{c,c}$ is the average temperature in the circular model. The sub-index $c$ refers to the case where the temperature is averaged inside $r\simeq0.025$.
The difference of having elliptical orbits instead of circular orbits in the cluster, gives as a result a decrease in the averaged temperature. That is, when the stars in the cluster have circular orbits, the energy produced from the colliding stellar winds is higher than in the elliptical case.

In the same way as in Equation \ref{ec:aries}, we define
$\aries_{sc,c}$ as
\begin{equation}
    \aries_{sc,c} = \left| 1 - \frac{T_{sc,c}}{T_{c,c}} \right| = 0.06  \mbox{ .}
    \label{ec:aries}
\end{equation}
In the latter equation $T_{sc,c}$ corresponds to  the average temperature value in the small circular orbit model. The behaviour of the small circular and the circular models is very similar.

In an analogous way we computed $\aries_{e,a} = 0.3$, $\aries_{sc,a} =0.7$,  $\aries_{e,b} = 0.17$, and  $\aries_{sc,b} = 0.25$. Where the sub-index $a$ and $b$ corresponds to the temperature averaged inside a radius $r\simeq0.05$ and $r\simeq0.1$, respectively. The net effect when considering elliptical orbits over circular orbits (for all radii analysed) is to decrease the average temperature. The net effect of reducing the size of the cluster radius a factor two, is to increase the temperature, a factor $1.25$ when considering the average temperature inside $R\simeq0.05$ pc, and a factor $1.7$ when considering the average temperature inside $R\simeq0.1$ pc.

In Figure~\ref{fig:density} we show the time evolution of the Y-midplane number density ($\mathrm{cm}^{-3}$) in the same layout as in Figure~\ref{fig:temp}. In panels (a) and (c) we show the case where the orbits are circular ($r_c=5.6$ and $~2.8$, respectively), and in panel (b) the case where the orbits are eccentric ($r_c=5.6$). We also show the radius of the cluster as a white circle.

In Figure~\ref{fig:density}  we can see that in both circular and eccentric orbits models a common cluster wind is formed, reaching a quasi stationary state somewhere between $100$ and $500~\mathrm{yr}$ of evolution.

With the results of the simulations (density and temperature stratifications)
it is possible to estimate the intrinsic X-ray emission from the cluster + SMBH
system. For this purpose we compute the X-ray emissivity in two energy bands: one
from $0.2$ to $2~\mathrm{Kev}$ (soft X-rays), and one from $2$ to $10~\mathrm{Kev}$
(hard X-rays). We use the {\sc chianti} atomic database and its software
\citep{dere97, dere19} and assume that the gas is in coronal equilibrium, and that
the emission is in the low-density regime (e.g. the emissivity  proportional to the
square of the density). The emission coefficient is then integrated taking the
y-coordinate as the line of sight. We show the resulting emission maps in Figure  \ref{fig:xray}.

We can see that there is an extended emission in the soft X-ray band that is more concentrated towards the center of the cluster but fills the entire domain. At the same time there is a significant emission in hard X-rays, which occurs mostly in the regions where the individual winds interact.

As the common wind forms, the overall X-ray luminosity increases with time and reaches a quasi-steady value after $\sim 500~\mathrm{yr}$.
The average soft X-ray intrinsic luminosity value after the initial transient is ${7.8\times 10^{34}~\mathrm{erg~s^{-1}}}$ for the model with circular orbits and $r_c=5.66$, ${6.9\times 10^{34}~\mathrm{erg~s^{-1}}}$ for the one with elliptical orbits and $r_c=5.6$, and ${9.2\times 10^{34}~\mathrm{erg~s^{-1}}}$ for the model with circular orbits and $r_c=2.8$. For the hard X-ray intrinsic luminosities the values are ${3.6\times 10^{34}~\mathrm{erg~s^{-1}}}$ (circular orbits with $r_c=5.6$), ${2.3\times 10^{34}~\mathrm{erg~s^{-1}}}$ (elliptical orbits with $r_c=5.6$), and  ${2.9\times 10^{34}~\mathrm{erg~s^{-1}}}$ (circular orbits with $r_c=2.8$).


\subsection{The analytic and numerical model compared}
\begin{figure}
 \centering
 \includegraphics[width=7.0cm]{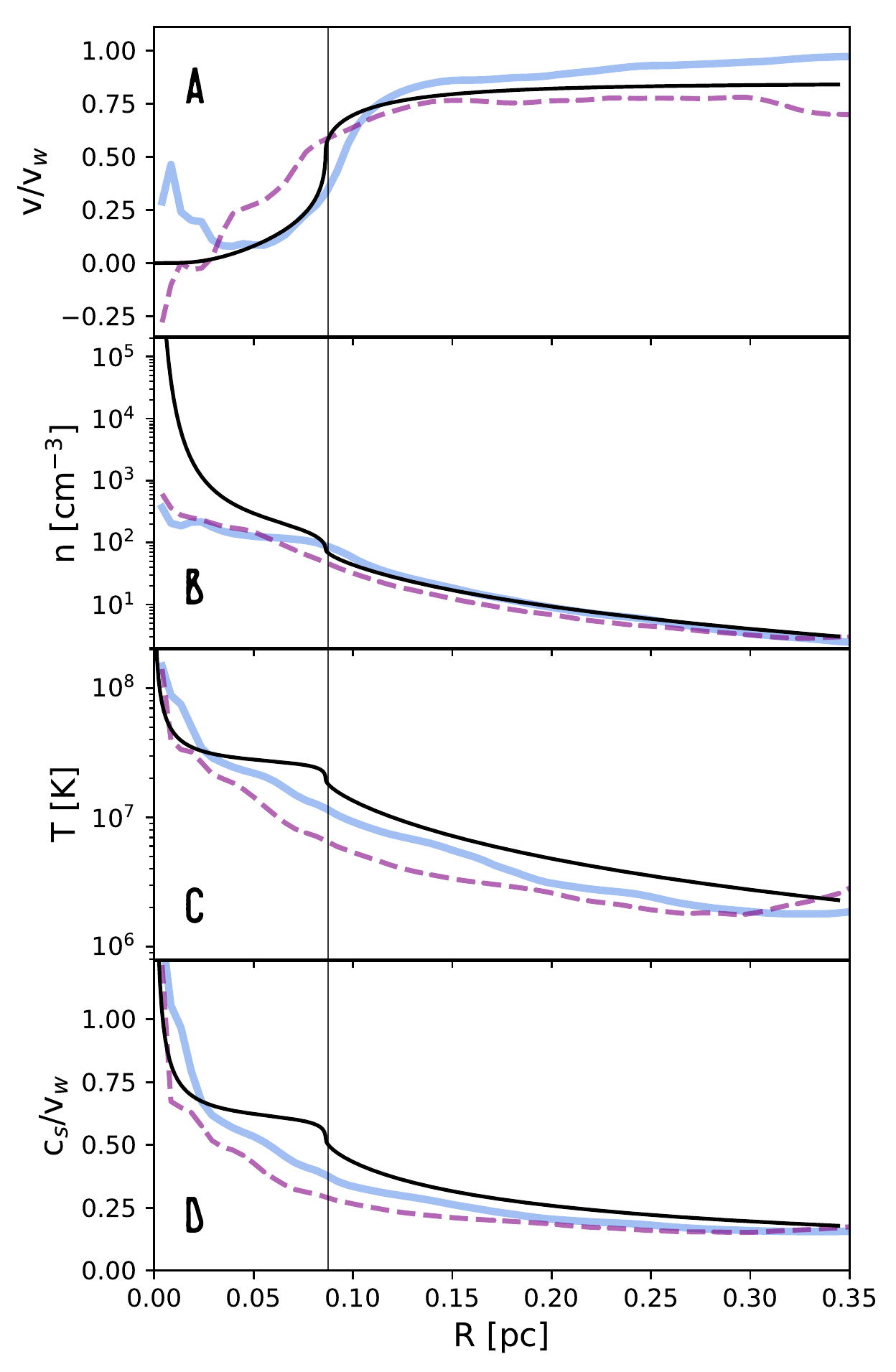}
 \caption{In this Figure we show at an integration time t=1 kyr the velocity normalized by the wind velocity ($1000~\mathrm{km\,s^{-1}}$, A), the number density (B), the temperature (C), and the sound speed normalised by the wind velocity (D) averaged over all directions, as a function of radius. The blue solid lines in all panels show the circular orbit case, the purple dashed lines shows the elliptical orbit case, and the black line show the analytical solution. The black vertical line shows the value of R=$R_{c}$, the star cluster radius.
 }
 \label{fig:compara}
 \end{figure}

As can be seen in Figure \ref{fig:compara}, the temperature reaches a stationary state after $\sim 0.5$ kyr. When the system has reached equilibrium, the numerical results can be compared with the analytical model described in Section \ref{sec:analytic}.

\begin{figure}
 \centering
 \includegraphics[width=7.0cm]{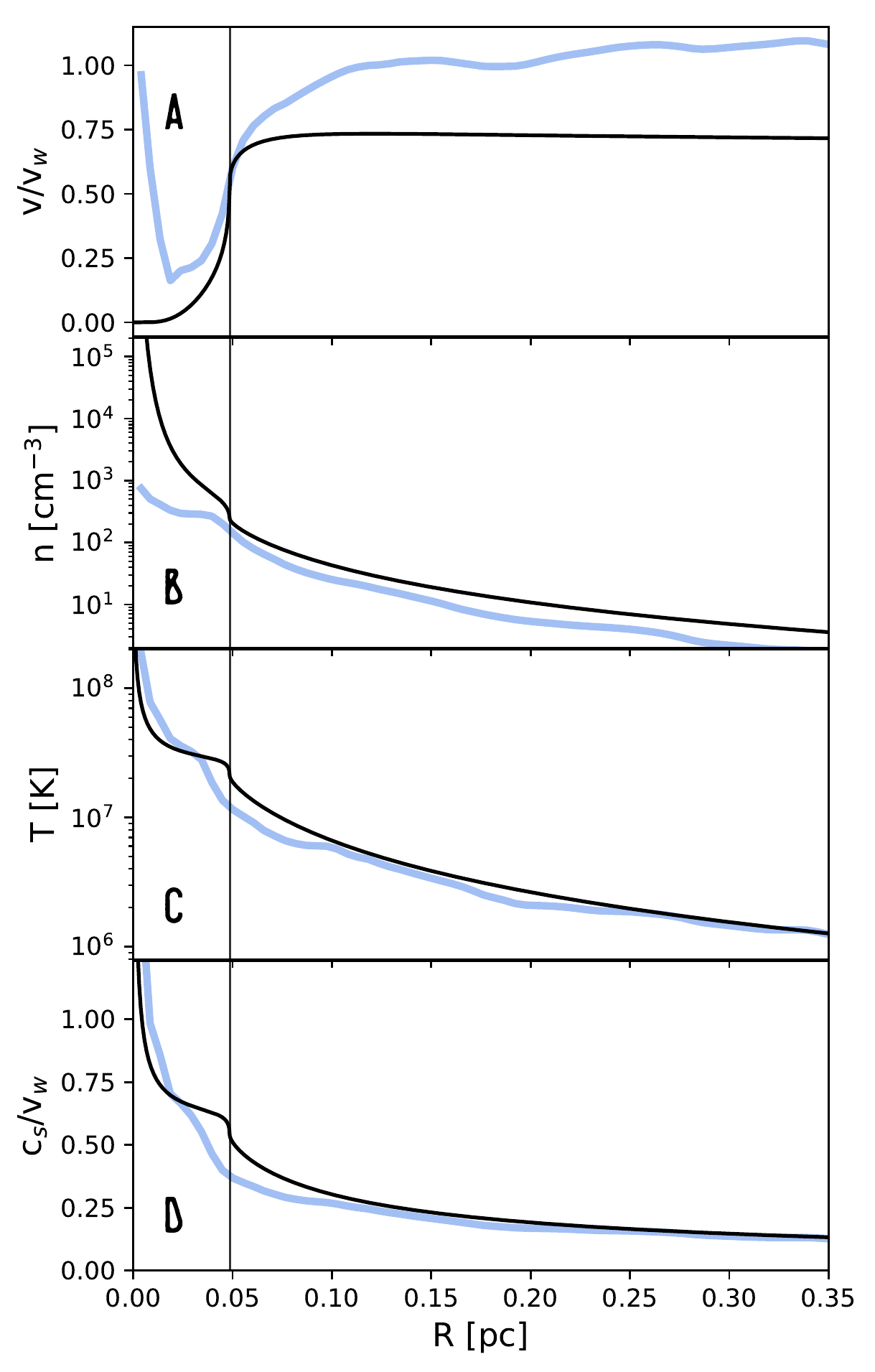}
 \caption{Same as Figure \ref{fig:compara} but for the model with circular orbits and $r_c=2.8$.
 }
  \label{fig:compara2}
 \end{figure}

Figure \ref{fig:compara} show the velocity (a), the density (b), the temperature (c), and the sound speed (d) as a function of the distance to the center of the star cluster (averaged over all directions) at the end of our simulation ($1$ kyr) for the two models with $r_c=5.6$. The solid blue line indicates the case where the stars are set in circular orbits, and the dashed purple line shows the results for the elliptical orbit case (all panels in Figure \ref{fig:compara} have the same color code). It is very clear from Figure \ref{fig:compara} that at this integration time, the net effect of stars orbiting in either circular or  elliptic orbits is almost indistinguishable. Only a small deviation is seen for the velocity (panel a) in the central regions of the stellar cluster.
The black lines in each panel of Figure \ref{fig:compara} show the analytical model derived in Section \ref{sec:analytic}.
One difference between the analytical model and the simulations is the lack of a sharp transition at $r_c$ in the latter. This can be attributed to the finite number of stars which make the boundary of the cluster not as sharply defined. In fact, for the case with elliptical orbits in which the radius of the cluster is slightly varying with time produces smoother radial profiles (see for instance the velocity profile).
In Figure \ref{fig:compara2} we show the comparison between the analytical model and the radial profiles resulting from the model with circular orbits and a smaller cluster radius ($r_c=2.8$). In this case we see also a satisfactory agreement between the analytical model and the simulation at distances $r > r_c$, except for the case of the velocity.

We obtain a very good agreement between the analytic and numerical approach. The agreement is specially good outside the cluster radius, which is shown as a vertical black lines in Figures \ref{fig:compara} and \ref{fig:compara2}.
Inside the cluster radius the agreement between the two calculations is quite good as well, except for the velocity profile (panel a).

The difference between the analytic cluster wind velocity
and the results from our simulation appears to be the
result of the fact that the mean outflow velocity
(computed from the numerical simulations) is strongly
affected by the velocities of the individual stellar
winds (while the analytic model only describes
a flow made of the combined winds from all of the stars).
This effect is clearly less important outside the
initial stellar cluster radius (where there are
basically no stars).

\subsection{Comparison with previous studies}
As we mentioned before, \cite{ressler:18} studied the accretion from strong stellar winds into the SMBH (Sgr A$^{*}$) in the the center of the Galaxy. They modeled the accretion  flow generated by 30 Wolf-Rayet stars orbiting Sgr A$^{*}$ (with an inner boundary of $r_{in}=6\times 10^{-5}$ pc). They adopted a mass for the SMBH of $M_{BH}=4.3\times10^{6}$~M$_{\odot}$. The mass loss rate and the velocities of each of the stellar winds were taken from the observational data of \cite{cuadra:08}.

\begin{figure}
 \centering
 \includegraphics[width=6.0cm]{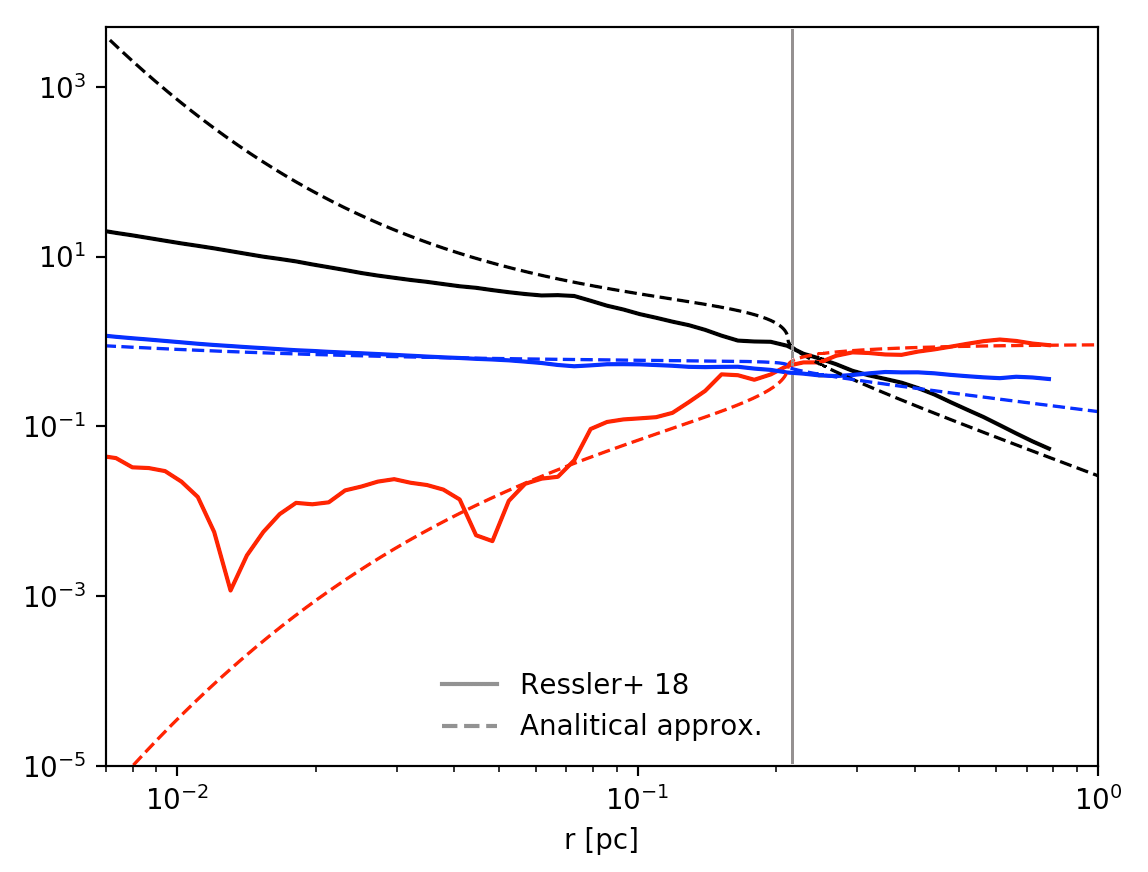}
 \caption{In this Figure we show the radial velocity in red (pc/kyr), the sound speed in blue (pc/kyr) and the density in black (M$_{\odot}$/pc$^{3}$). The solid lines show the numerical results of \cite{ressler:18} (their Figure 18). The dashed lines show our analytical model (derived from Section \ref{sec:analytic}). The gray vertical line shows the cluster radius $R_c = 0.22$ pc (which corresponds to $r_c = 11.9$).}
 \label{fig:ressler}
 \end{figure}

We calculated the average mass loss rate and velocity from \cite{cuadra:08}, and set the mass of the SMBH to $4.3\times10^{6}$~M$_{\odot}$. With those parameters we could build an analytical approximation following Section \ref{sec:analytic}. In Figure \ref{fig:ressler} we show the results of \cite{ressler:18} (their Figure $18$) and compare it to our analytical results. The value of the dimensionless cluster radius for the analytical model is  $r_c \simeq11.9$, which corresponds to
$R_c\simeq 0.22~\mathrm{pc}$, and it is indicated by the vertical gray line in Figure \ref{fig:ressler}. The solid lines in Figure \ref{fig:ressler} correspond to \citeauthor{ressler:18}'s \citeyear{ressler:18} data and the dashed line correspond to our analytic results. In black we show the density with units $M_{\odot}/\mathrm{pc}^{3}$, the radial velocity is shown in red (with units $\mathrm{pc/kyr}$), and the sound speed is shown in blue (with units $\mathrm{pc/kyr}$), all of which are shown as a function of the distance to the center of Sgr A$^{*}$.

It is very encouraging to obtain a good agreement between \citeauthor{ressler:18}'s \citeyear{ressler:18} results and our analytical approach. This is a second test which validates the analytical approach and guarantees that this approach can be used to infer some of the properties of the flow generated from strong stellar winds orbiting a massive compact object.

\section{Conclusions}
\label{sec:conc}

In this work we studied for the first time an analytical approximation of
a system consisting of a cluster of stars (where all stars present strong winds), orbiting around a massive particle (mimicking a SMBH). In order to validate the analytical model, we run 3D HD numerical simulations of a cluster of stars where each of the stars have a strong associated wind, and are orbiting a massive object in circular and elliptical orbits.

We set all stars with the same parameters of velocity, density, temperature and wind mass-loss rate. We studied three different cases for the orbits of the stars: the first two cases with stars in circular orbits around a SMBH with a mass of $4\times10^{6}$ M$_{\odot}$ and different cluster radius, and a third case where the star orbits are elliptical. We let our numerical simulation run from the initial conditions (described in Section \ref{sec:IC}) to an integration time of $1$ kyr.

We found that after the system has reached a quasi-stationary state (after $\sim 0.5$ kyr) both orbital cases (circular and elliptical) behave virtually the same. We only observe a small deviation between both orbital cases in the velocity profile. In the top panel of Figure \ref{fig:compara} it can be seen that the magnitude of the velocity profile inside the cluster is greater for the elliptical case (purple dashed line). We can explain this discrepancy since the elliptical orbital velocities are greater when the stars approach the central region of the cluster where the SMBH is located, thus having overall a greater contribution to the average velocity. Still the difference between the radial velocity profiles is rather small.
A very important conclusion in this work is that the temperature and density profile (and thus the X-ray emission) from the flow generated as a consequence of shocks of the stellar winds orbiting a central SMBH, depend on the net energy injected via the mass loss of the stars. The total mass loss of the stars in our simulation is $\sim10^{-3}$~M$_{\odot}$/yr.

We did not include the effect of magnetic fields in the stellar winds in our numerical calculations since \cite{ressler:20} found that the effect of including magnetized stellar winds is of minor importance.

In general, we found a good agreement between the analytical model and the numerical simulations. The fact that such a complex system of a flow generated from the shocks of strong winds interacting with a SMBH, can be summarized in a simple ``mass loaded flow" model is quite remarkable. This result allows us to make a simple and quick calculation of the properties of these systems, even if one does not have all the orbital stellar information.

Moreover, \cite{ressler:18} performed 3D HD simulations where they included the most up-to-date positions, speeds and mass-loss rates of 30 (Wolf-Rayet) stars located in the central parsec of the Galactic center. They include the accretion to the BH, and the cooling from a tabulated version of the exact collisional ionization equilibrium cooling function. We also obtain an excellent agreement from our analytical calculations when compared with \citeauthor{ressler:18}'s \citeyear{ressler:18} numerical simulations.

To summarize, we have derived an analytic model for
the combined wind from a cluster of stars (with
strong stellar winds) with a central BH. We
find that this model produces predictions of the
cluster wind that are in good agreement with
the results obtained from numerical simulations.
Therefore, the analytic model is completely
appropriate for obtaining predictions of the
characteristics of the cluster wind, particularly for
radii $r \gtrsim r_c$.

%
 %

\acknowledgments
We thank the anonymous referee for very kind and useful comments that improved the presentation of this paper.
V.L. thanks Sean Ressler for making his data available.
V.L. gratefully acknowledges support from the \mbox{CONACyT} Research Fellowships program.
A.R., J.C., and A. E. acknowledge support from DGAPA-PAPIIT (UNAM) grant BG100218. We also thank Enrique Palacios Boneta (HPC system manager, ICN-UNAM) for maintaining our cluster, where the simulations were carried out.
{\sc Chianti} is a collaborative project involving George Mason University, the University of Michigan (USA), University of Cambridge (UK) and NASA Goddard Space Flight Center (USA).




\clearpage


\begin{thebibliography}{99}
%
\bibitem[\protect\citeauthoryear{Abuter et al.}{2018}]{abu18}
Gravity Collaboration; Auter, R., Amorim, A., Anugu, N., et al. 2018, A\&A, 615, L15

\bibitem[\protect\citeauthoryear{Baganoff et al.}{2003}]{bag03}
Baganoff, F. K., Maeda, Y., Morris, M., et al. 2003, ApJ, 591, 891

\bibitem[\protect\citeauthoryear{Cant\'o et al.}{2000}]{can00}
Cant\'o, J., Raga, A. C., Rodr\'\i guez, L. F. 2000, ApJ, 536, 896

\bibitem[\protect\citeauthoryear{Cuadra et al.}{2005}]{cuadra:05}
Cuadra, J., Nayakshin, S., Springel, V., et al.\ 2005, MNRAS, 360, L55

\bibitem[\protect\citeauthoryear{Cuadra et al.}{2006}]{cuadra:06}
Cuadra, J., Nayakshin, S., Springel, V., et al.\ 2006, MNRAS, 366, 358

\bibitem[\protect\citeauthoryear{Cuadra, Nayakshin \& Martins}{2008}]{cuadra:08}
Cuadra, J., Nayakshin, S., \& Martins, F.\ 2008, MNRAS, 383, 458

\bibitem[\protect\citeauthoryear{Castellanos-Ram\'irez et al.}{2015}]{cas:15}
Castellanos-Ram\'irez, A., Rodr\'iguez-Gonz\'alez, A.,Esquivel, A., et al. 2015, MNRAS, 450, 2799

\bibitem[\protect\citeauthoryear{Chevalier \& Clegg}{1985}]{chevalier:85}
Chevalier, R. A., Clegg, A. W., 1985, Nature, 317, 44

\bibitem[\protect\citeauthoryear{Dere et al.}{1997}]{dere97}
Dere, K.~P., Landi, E., Mason, H.~E., et al.\ 1997, A\&AS, 125, 149

\bibitem[\protect\citeauthoryear{Dere et al.}{2019}]{dere19}
Dere, K.~P., Del Zanna, G., Young, P.~R., et al.\ 2019, ApJS, 241, 22

\bibitem[\protect\citeauthoryear{Esquivel et al.}{2009}]{esquivel:09}
Esquivel, A.,  Raga A. C., Cant\'o J., Rodr\'iguez-Gonz\'alez A., 2009, A\&A, 507, 855

\bibitem[\protect\citeauthoryear{Esquivel \& Raga}{2013}]{esquivel:13}
Esquivel, A. \& Raga A. C., 2013, ApJ, 779, 111

\bibitem[\protect\citeauthoryear{Falle et al.}{2012}]{falle02}
Falle, S. A. E. G., Coker, R. F., Pittard, J.M., Dyson, J. E., Hartquist, T. W. 2002, MNRAS, 329, 670

\bibitem[\protect\citeauthoryear{Genzel et al.}{2003}]{gen03}
Genzel, R., Sch\"odel, R., Ott, T., et al. 2003, ApJ, 599, 812

\bibitem[\protect\citeauthoryear{Ghez et al.}{2005}]{ghez:05}
Ghez A. M., Salim S., Hornstein S. D., et al. 2005, ApJ, 620, 744


\bibitem[\protect\citeauthoryear{Hueyotl-Zahuantitla et al.}{2010}]{hueyotl:10}
Hueyotl-Zahuantitla, F., Palous, J., W\"unsch, R.
et al. 2010, ApJ, 766, 92

\bibitem[\protect\citeauthoryear{Lora et al.}{2009}]{lora:09}
Lora, V., S\'anchez-Salcedo, F. J., Raga, A. C. \& Esquivel, A., 2009, ApJL, 699, L113

\bibitem[\protect\citeauthoryear{L\"utzgendorf et al.}{2016}]{lutzgendorf:16}
L{\"u}tzgendorf, N., van der Helm, E., Pelupessy, F.~I., et al.\ 2016, MNRAS, 456, 3645

\bibitem[\protect\citeauthoryear{Martins et al.}{2007}]{martins:07}
Martins F., Genzel R., Hillier D. J., Eisenhauer F., Paumard T.,Gillessen S., Ott T., Trippe S., 2007, A \& A, 468, 233

\bibitem[\protect\citeauthoryear{Palou\u s et al.}{2013}]{pal13}
Palou\u s, J., W\"unsch, R., Mart\'\i nez-Gonz\'alez, S., et al. 2013, ApJ, 772, 128

\bibitem[\protect\citeauthoryear{Quataert}{2004}]{qua:04}
  Quataert, E. 2004, ApJ, 613, 344

\bibitem[\protect\citeauthoryear{Raga et al.}{2001}]{rag01}
Raga, A. C., Vel\'azquez, P. F., Cant\'o, J., Masciadri, E., Rodr\'\i guez, L. F. 2001, ApJ, 559, L33

\bibitem[\protect\citeauthoryear{Ressler, Quataert \& Stone}{2018}]{ressler:18}
Ressler, S. M., Quataert, E. \& Stone, J. M. 2018, MNRAS, 478, 3544

\bibitem[\protect\citeauthoryear{Ressler, Quataert \& Stone}{2020}]{ressler:20}
Ressler, S. M., Quataert, E. \& Stone, J. M. 2020, MNRAS, 492, 3272

\bibitem[\protect\citeauthoryear{Rockefeller et al.}{2004}]{rockefeller:04}
Rockefeller, G., Fryer, C., Melia, F., Warren, M. S., 2004, ApJ, 604, 662

\bibitem[\protect\citeauthoryear{Rodr\'\i guez-Gonz\'alez et al.}{2007}]{rodg07}
Rodr\'\i guez-Gonz\'alez, A., Cant\'o, J., Esquivel, A., Raga, A. C., Vel\'azquez, P. F. 2007, MNRAS, 380, 1198

\bibitem[\protect\citeauthoryear{Rodr\'\i guez-Gonz\'alez et al.}{2008}]{rodg08}
Rodr\'\i guez-Gonz\'alez, A., Esquivel, A.,Raga, A. C., Cant\'o, J. 2008, ApJ, 684, 1384

\bibitem[\protect\citeauthoryear{Rodr\'\i guez-Ram\'irez et al.}{2014}]{rodr14}
Rodr\'\i guez-Ram\'\i rez, J. C., Raga, A. C., Vel\'azquez, P. F., Rodr\'\i guez-Gonz\'alez, A.,
Toledo-Roy, J. C. 2014, MNRAS, 445, 1023

\bibitem[\protect\citeauthoryear{Shcherbakov \& Baganoff}{2010}]{shc:10}
Shcherbakov, R. V., Baganoff, F. K. 2010, ApJ, 716, 504

\bibitem[\protect\citeauthoryear{Silich et al.}{2004}]{silich:04}
Silich, S., Tenorio-Tagle, G. \& Rodr\'\i guez-Gonz\'alez, A. 2004, ApJ, 610, 226

\bibitem[\protect\citeauthoryear{Silich et al.}{2008}]{silich:08}
Silich, S., Tenorio-Tagle, G., Hueyotl-Zahuantitla, F.
2008, ApJ, 686, 172

\bibitem[\protect\citeauthoryear{Toro E. F.}{1999}]{toro:99}
Toro E. F., 1999, Riemann Solvers and Numerical Methods for Fluid Dynamics. Springer-Verlag, Berlin

\bibitem[\protect\citeauthoryear{Wang et al.}{2013}]{wang13}
Wang, Q. D., Nowak, M. A., Markoff, S. B., et al. 2013, Sci, 341, 981

\bibitem[\protect\citeauthoryear{Yalinevich et al.}{2018}]{yal:18}
Yalinevich, A., Sari, R., Generosov, A., Stone, N. C., Metzgter, B. D.
2018, MNRAS, 479, 4778Y

\bibitem[\protect\citeauthoryear{Yusef-Zadeh et al.}{2016}]{yusef:16}
Yusef-Zadeh, F., Wardle, M., Sch\"odel, R., et al. 2016, ApJ, 819, 60


\end{thebibliography}
\end{document}